\documentclass{jfm}
\usepackage[english]{babel}
\usepackage{graphicx}
\usepackage{dcolumn}
\usepackage[outline]{contour}
\usepackage[utf8]{inputenc}
\usepackage[T1]{fontenc}

\usepackage{amsmath,amsfonts,amssymb,amsbsy}

\usepackage{hyperref}
\hypersetup{
	pdfborder={0 0 0}
	}
\usepackage{xcolor}
\usepackage{cleveref}
\usepackage{graphicx}
\usepackage{physics}
\usepackage{tikz}
\usetikzlibrary{chains}
\usetikzlibrary{patterns}
\usepackage{lmodern}
\usepackage{lipsum}
\usepackage{csquotes}
\DeclareRobustCommand\full  {\tikz[baseline=-0.6ex]\draw[thick] (0,0)--(0.5,0);}

\DeclareRobustCommand\dotted{\tikz[baseline=-0.6ex]\draw[thick,dotted] (0,0)--(0.54,0);}
\DeclareRobustCommand\dashed{\tikz[baseline=-0.6ex]\draw[thick,dashed] (0,0)--(0.54,0);}
\DeclareRobustCommand\Rdashed{\tikz[baseline=-0.6ex]\draw[,red,thick,dashed] (0,0)--(0.54,0);}
\usepackage{adjustbox}
\usepackage{multirow}
\usepackage{subcaption}
\usepackage{MnSymbol}
\usetikzlibrary{positioning,3d}
\usetikzlibrary{graphs,automata,matrix}
\usepackage{float}
\usepackage{pgfplots}
\usepgfplotslibrary{fillbetween,groupplots}
\usepackage{bm}

\pgfplotsset{compat=1.15
 ,colormap={parula}{
rgb255=(53,42,135)
rgb255=(15,92,221)
rgb255=(18,125,216)
rgb255=(7,156,207)
rgb255=(21,177,180)
rgb255=(89,189,140)
rgb255=(165,190,107)
rgb255=(225,185,82)
rgb255=(252,206,46)
rgb255=(249,251,14)
        }}
        \shorttitle{DNS of pseudo-random roughness in minimal channels}
\shortauthor{J. Yang, A. Stroh, D. Chung and P. Forooghi}
    \title{Direct numerical simulation-based characterization of pseudo-random roughness in minimal channels}

\author{Jiasheng Yang\aff{1}, Alexander Stroh\aff{1}, Daniel Chung\aff{2}\and Pourya Forooghi\aff{3}\corresp{\email{forooghi@mpe.au.dk}}}

\affiliation{\aff{1}Institute of Fluid Mechanics, Karlsruhe Institute of Technology, Karlsruhe, Germany
\aff{2}Department of Mechanical Engineering, University of Melbourne, Victoria 3010, Australia\aff{3}Department of Mechanical and Production Engineering, Aarhus University, Aarhus, Denmark}

\begin{document}


\maketitle

\begin{abstract}
    Direct numerical simulations (DNS) are used to systematically investigate the applicability of the minimal channel approach~\citep{chung_chan_macdonald_hutchins_ooi_2015} for the characterization of roughness-induced drag on irregular rough surfaces. Roughness is generated mathematically using a random algorithm, in which the power spectrum (PS) and probability density function (PDF) of surface height function can be prescribed. 12 different combinations of PS and PDF are examined and both transitionally and fully rough regimes are investigated (roughness height varies in the range $k^+$ = 25 -- 100).
    It is demonstrated that both the roughness function ($\Delta U^+$) and the zero-plane displacement can be predicted with $\pm5\%$ accuracy using DNS in properly sized minimal channels. Notably, when reducing the domain size, the predictions remain accurate as long as 90\% of the roughness height variance is retained. Additionally, examining the results obtained from different random realizations of roughness shows that a fixed combination of PDF and PS leads to a nearly unique $\Delta U^+$ for deterministically different surface topographies.
    In addition to the global flow properties, the distribution of time-averaged surface force exerted by the roughness onto the fluid is calculated and compared for different cases. 
    It is shown that patterns of surface force distribution over irregular roughness can be well captured when the sheltering effect is taken into account. This is made possible by applying the sheltering model of \citet{yang16} to each specific roughness topography.
    Furthermore, an analysis of the coherence function between the roughness height and the surface force distributions reveals that the coherence drops at larger streamwise wavelengths, which can be an indication that very large horizontal scales contribute less to the skin-friction drag. 
\end{abstract}
\begin{keywords}
	turbulent simulation, roughness, minimal channel
\end{keywords}
\section{Introduction}
Turbulent flows bounded by rough walls are abundant in both Nature - (e.g. fluvial flows \citep{mazzuoli17} and wind flow over vegetation~\citep{SHAW198251} and urban canopies~\citep{coceal04,yang16}) - and industry - e.g. degraded gas turbine blades \citep{bons01}, bio-fouled ship hulls \citep{hutchins16}, iced surfaces in aero-engines~\citep{Juan2019} and deposited surfaces inside combustion chambers~\citep{FOROOGHI201883}. 
Systematic study of roughness effects on skin friction dates back to the pioneering works of ~\citet{Nikuradse1933} and ~\citet{schlichting1936experimentelle}. 
Flow-related roughness is usually classified into two types, k-type and d-type roughness. 
For k-type roughness the flow response depends directly on the physical scale of the roughness height, while for d-type roughness it is insensitive to the roughness height scale, but rather determined by the outer length scales, e.g. the pipe diameter $d$ \citep{perry_schofield_joubert_1969,jimenez04}.
The scope of the present work is limited to three-dimensional (3-D) irregular rough surfaces, in which k-type behaviour is relevant.

In industry, the Moody diagram \citep{Moody1944} has been considered as a standard method to calculate the skin friction of a rough surface.
The Moody diagram relates the friction factor to the roughness height $\epsilon$, which is linked to the equivalent sand-grain roughness $k_s$. This quantity, which is also an input to many low-fidelity turbulence models for rough walls \citep{Suga06,Brereton18}, is not known \textit{a priori} for any given irregular rough surface.
Hence, for any new roughness topography, $k_s$ needs to be determined using a laboratory or high-fidelity numerical experiments or estimated based on roughness correlations derived from such experiments.
The problem of predicting the roughness-induced friction drag based merely on the knowledge of the roughness topography has received extensive attention in the past, and a variety of roughness correlations have been developed in different industrial contexts \citep{waigh98,macdonald00,Rij,BONSRA,flack2010,chan2015,10.1115/1.4037280,Thakkar2017,Flack2020}. 
In these roughness correlations, the topography of the rough surface is often represented by statistical measures of the roughness height map $k(x,z)$, with $k$ being the surface height as a function of horizontal coordinates $x$ and $z$. 
Some widely discussed statistical measures in this context are summarized in the recent review of the topic by~\citet{Chung2021annrev}, for instance the skewness $Sk$ \citep{flack2010,10.1115/1.4037280}, effective slope ES~\citep{napoli_armenio_demarchis_2008,chan2015} and density parameter $\Lambda_s$~\citep{Sigal,Rij}. 
Despite extensive work in the past, a universal correlation with the ability to accurately predict the drag of a generic rough surface remains elusive \citep{flack18}. Arguably, the development of such a correlation requires a large amount of data from realistic roughness samples. 
However, the generation of an appropriate database has been hindered mainly due to two factors: the formidable cost associated with many numerical or laboratory experiments, and the relative scarceness of realistic roughness maps combined with the lack of ability to systematically vary their properties. 

A considerable portion of data in the literature deals with regular roughness - often generated by the distribution of similar geometric elements. Examples of the geometries studied include cubes~\citep{Orlandi06,leonardi_castro_2010}, spheres~\citep{mazzuoli17}, pyramids~\citep{Schultz2009}, LEGO bricks~\citep{placidi_ganapathisubramani_2015}, ellipsoidal egg-carton shape~\citep{Bhaganagar2008}, and sinusoidal roughness~\citep{chan2015,chan_macdonald_chung_hutchins_ooi_2018}. In comparison, investigations based on realistic rough surfaces are less frequent and include a much lower number of cases. Notably, ~\citet{Thakkar2017} utilized direct numerical simulation (DNS) to study the effect of roughness topography on flow statistics for 17 industrially relevant irregular surfaces and proposed roughness correlations for the transitionally rough regime. Other examples of realistic roughness studies in the framework of wall-bounded turbulence include ~\citet{cardillo_chen_araya_newman_jansen_castillo_2013,yuan14,BUSSE2015129,busse_thakkar_sandham_2017,FOROOGHI201883,Yuan2018,Jouybari19,Mangavelli21}.

In recent years, mathematically generated surfaces have received an increased amount of attention as a means to systematically study realistic irregular roughness. 
Many of these roughness generation approaches rely on random superposition of discrete geometric elements  ~\citep{scotti2006,chau2012,Forooghiheattransfer,kuwata_kawaguchi_2019} or Fourier modes~\citep{anderson_meneveau_2011,demarchis20}.
Some authors \citep{BARROS20181,jelly19} opt for approaches based on the linear combination of random numbers in an attempt to study surfaces that resemble realistic roughness as closely as possible. These methods can produce a prescribed power spectrum (PS) with a Gaussian probability density function (PDF) for the generated roughness. Motivated by the relevance of non-Gaussian roughness in industry, recently \citet{Flack2020} employed a modified version of such methods to study non-Gaussian roughness with a certain choice of PDFs.
In the present paper, we adopt an alternative roughness generation method proposed by~\cite{PEREZRAFOLS2019591}. This method is deemed advantageous in producing surrogates of realistic roughness since it provides absolute flexibility to prescribe any desired combination of PDF and PS as well as more robustness compared to algorithms based on translations of the Pearson's or Johnson's types used by previous authors. 
We refer to the roughness samples generated by this method as `pseudo-random' roughness in the sense that the topography is random but its statistical properties are prescribed.

In recent years, DNS has been the pacing approach in studying the effect of roughness topography on friction drag. 
Standard DNS, however, involves resolving the entire spectrum of turbulent length scales ranging from large geometrical scales to the small viscous scale, which is computationally costly. 
To tackle this problem,~\citet{chung_chan_macdonald_hutchins_ooi_2015} and \citet{MacDonald_2016} employed the idea of DNS in minimal span channels~\citep{jimenez_moin_1991,Flores2010} for prediction of roughness-induced drag over a regular sinusoidal roughness in a channel. The central idea followed by these authors is that the amount of downward shift in the inner-scaled velocity profile $\Delta U^+$ is the determining factor in the prediction of drag. 
These authors showed that, thanks to outer layer similarity of wall bounded turbulence~\citep{Townsend}, this key quantity can be accurately predicted by minimal rough channels. 
This can be achieved as long as an adequately large range of near-wall structures are accommodated in the simulation domain. 
The same group of authors further developed the idea for channels with minimal streamwise extent~\citep{macdonald_chung_hutchins_chan_ooi_garcia-mayoral_2017}, high-aspect-ratio transverse bars~\citep{macdonald_ooi_garcia-mayoral_hutchins_chung_2018} and also for passive scalar calculations ~\citep{macdonald_hutchins_chung_2019}. 
These efforts established 
the following criteria for the size of a minimal channel based on simulations with 3-D sinusoidal roughness:
 \begin{equation}
 L_z^+\geq \text{max}(100,\frac{\Tilde{k}^+}{0.4},\lambda_{\sin}^+)~,L_x^+\geq \text{max}(1000,3L_z^+,\lambda_{\sin}^+)~.
 \label{MINI2}
 \end{equation}
Here $L_z$ and $L_x$ are the spanwise and streamwise extents of the minimal channel, respectively, $\lambda_{\sin}$ is the sinusoid wavelength of roughness; $\Tilde{k}$ is the characteristic roughness height (here sinusoid amplitude) and the plus superscript indicates viscous scaling. 
While the aforementioned studies showed the potential of minimal channels in the determination of roughness-induced drag, a formal extension of this concept to random, irregular roughness is yet to be made. 
Recently,~\citet{alves20} reported predictions of flow over a realistic roughness combining minimal channels with a novel hybrid DNS/URANS model. These authors highlighted the need for careful investigation of minimal channel concept for realistic roughness. 
Although some previous researchers, e.g. ~\cite{jouybari_2021,Pargal21}, have already applied the concept successfully for random surfaces, no systematic verification of minimal-channel approach for irregular roughness of random nature has been reported in the past. 

In  view of the above, the present work aims to provide a systematic proof for the validity of minimal-channel approach in irregular roughness of random nature. Here `systematic' refers to covering a wide range of roughness topographies (12 different PDF/PS combinations) and  both transitional and fully roughness regimes. The study aims to answer two key questions. The first is whether it is adequate for roughness in the minimal channel to be similar to the original roughness merely in a `statistical' sense. This, in turn, relates to the more fundamental question about the minimum amount of statistical information needed to uniquely predict the roughness-induced drag. The second question is whether the rules set by equation \ref{MINI2} can be modified or relaxed so that they are applicable to any type of roughness. Here the critical issue is that a realistic roughness may contain very large horizontal wavelengths, making the minimal channel approach futile if the original rules are to be met strictly. We compare the results from minimal- and full-channel DNS in section \ref{sec:EMC}. To shed more light on the latter question, in section \ref{IBMF} we also study in detail the local distribution of drag force on a rough surface to better understand the contribution of different horizontal scales in drag generation. As a final point, we use the generated data to assess a number of widely used roughness correlations in section \ref{results}. A detailed description of our methodology and a summary of findings are presented in sections \ref{Numerical} and \ref{Conclusion}, respectively.


\section{Numerical methodology}
\label{Numerical}
\subsection{Pseudo-random roughness generation}
\label{sec:gen}
As mentioned in the introduction, the roughness generation method proposed by~\citet{PEREZRAFOLS2019591} is adopted in the present methodology. 
In this method both the wall-parallel and the wall-normal statistical properties of the roughness can be adjusted.
Here wall-parallel properties refer to the PS of the roughness structures and wall-normal properties refer to the PDF of the surface height. 
The roughness map is represented by a discrete elevation distribution on a two-dimensional (2-D) Cartesian grid. 
The generation algorithm used in the present work takes the target PDF and PS as inputs. 
Transformations between the physical space and spectral space are done by discrete fast Fourier transform.
Initially a roughness map $k_\text{PDF}^0$ is generated which has the prescribed PDF but not necessarily the prescribed PS. 
{This initial map is then corrected in the Fourier space according to the prescribed PS, which is represented by $\hat{k}_{\text{PS}}^{i}$: 
    \begin{equation}
        \hat{k}_{\text{PS}}^{i+1}=\hat{k}_{\text{PDF}}^{i}\frac{|\hat{k}_{\text{PS}}^{i}|}{|\hat{k}_{\text{PDF}}^{i}|}.
    \end{equation}
    where $i$ indicates the iteration of the generation process. The output of this stage, $k_{\text{PS}}^{i+1}$, has the desired PS but not necessarily the prescribed PDF. In the present notation, subscripts PDF and PS indicate that the roughness field has the desired PDF or PS, respectively. The hat indicates the Fourier transform.
    After which $k_{\text{PDF}}^{i+1}$ is updated by correcting the PDF of $k_{\text{PS}}^{i+1}$ by rank ordering.} This correction process continues for $n$ iterations until both  $k_\text{PDF}^n$ and $k_{\text{PS}}^n$ converge to a height map with the target PDF and PS within a predetermined error.  
For more details on the the generation algorithm, readers are referred to~\citet{PEREZRAFOLS2019591}.

\subsection{Direct numerical simulation}
\label{sec:DNS}
\begin{figure}
\centering
   \includegraphics[width=10cm]{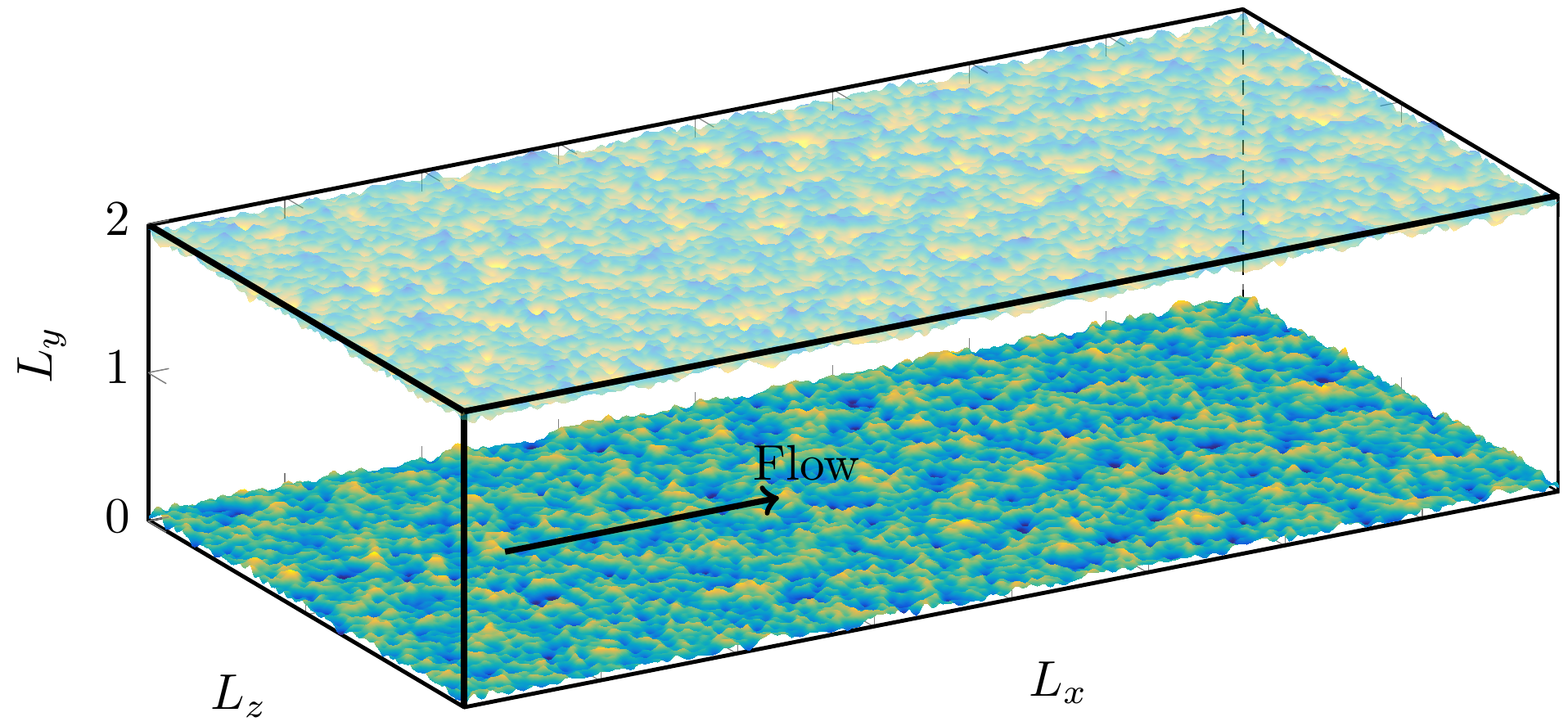}
    \caption{Schematic representation of the simulation domain with an example pseudo-realistic surface mounted. Normalization of lengths with $H$ is applied in this figure.}
    \label{fig:domain}
\end{figure}
A number of DNS have been carried out in fully developed turbulent plane channels, in which the flow is driven by a constant pressure gradient.
A representation of the simulation domain is shown in figure~\ref{fig:domain}, where $x$, $y$ and $z$ denote the streamwise, wall-normal and spanwise directions with respective velocity components $u$, $v$ and $w$.
The roughness elements are mounted on both the upper wall and the lower wall.
The channel half-height, which is the distance between the deepest trough in the roughness and the centre-plane of the channel, is labelled as $H$, and will be used to normalize the geometrical scales.
The incompressible Navier-Stokes equations are solved using the pseudo-spectral solver SIMSON~\citep{Chevalier}, where wall-parallel directions are discretized in Fourier space, while in the wall-normal direction Chebyshev discretization is employed.
The immersed Boundary Method (IBM) based on~\citet{goldstein93} is used to impose the no-slip boundary condition on the roughness by introducing an external volume force field directly to the Navier-Stokes equation.
The presently used IBM implementation has been validated and used in previous studies~\citep{PhysRevFluidsPourya,Vanderwel19,stroh_schaefer_frohnapfel_forooghi_2020}.\\
The Navier-Stokes equationcan be written as
       \begin{equation}
       \bnabla \bcdot \textbf{u}=0,
        \end{equation}
\begin{equation}
    \frac{\partial \textbf{u}}{\partial t}+\bnabla\bcdot(\textbf{uu})=-\frac{1}{\rho}\bnabla p+\nu\nabla^2\textbf{u}-\frac{1}{\rho}P_x{\mathbf{e_x}}+\textbf{f}_\text{IBM},
\end{equation}
where $\textbf{u}=(u,v,w)^\intercal$ is the velocity vector and $P_x$ is the mean pressure gradient in the flow direction added as a constant and uniform source term to the momentum equation to drive the flow in the channel. 
Here $p$, $\mathbf{e_x}$, $\rho$, $\nu$ and $\textbf{f}_\text{IBM}$ are pressure fluctuation, streamwise basis vector, density, kinematic viscosity and external body-force term due to IBM, respectively.
Periodic boundary conditions are applied in the streamwise and spanwise directions. 
A no-slip boundary condition is applied on the rough walls.
The friction Reynolds number is defined as Re$_\tau=u_\tau(H-k_{\text{md}})/\nu$, where $u_\tau=\sqrt{\tau_w/\rho}$ and $\tau_w=-P_x(H-k_{\text{md}})$ are the friction velocity and wall shear stress, respectively. The meltdown height, denoted by $k_{\text{md}}$, is the mean roughness height measured from the deepest trough. Note that we use $(H-k_{\text{md}})$ which is the mean half-cross-sectional area divided by the channel width -- or arguably the effective channel half-height -- as the reference length scale in these definitions. In total, four different values of Re$_\tau$ in the range of 250-1000 are simulated in the present work in order to be able to cover different regimes. The simulations designed to study the effect of roughness topography are, however, performed mainly at Re$_\tau=500$.

The simulation domain is discretized in an equidistant grid in wall-parallel directions, while in the wall-normal direction cosine stretching based on Chebyshev node distribution is applied.
The selection of grid size must take into consideration both flow and roughness length scales.
As reflected by ~\citet{BUSSE2015129}, each roughness wavelength should be represented by multiple computational cells.
Since we prescribe the PS in the roughness generation approach, the range of present roughness wavelengths can be prescribed.
Here the smallest roughness wavelength, labelled as $\lambda_1$, is the crucial quantity for horizontal grid resolution.
Therefore, in view of the present computational capacity, $\lambda_1=0.08H$ is prescribed for the following simulations.
A mesh independence study is carried out, from which the smallest roughness wavelength being resolved by 8-10 cells in each direction is found to be adequate to obtain the converged double-averaged velocity profile.
Overall, the grid size $\Delta^+<5$ in wall-parallel directions is proven to be appropriate through the mesh independence test.
In wall-normal directions, cosine stretching mesh is adopted for the Chebychev discretization.
It is also checked through the mesh independence test that, for present types of rough surfaces, a vertical cell number of 401 is sufficient in delivering a converged result at the highest Re$_\tau\approx1000$, thanks to the overresolving of the roughness structure by the cosine stretching grid near the wall.

\subsection{Description of cases}
\label{sec:cases}
\begin{figure}
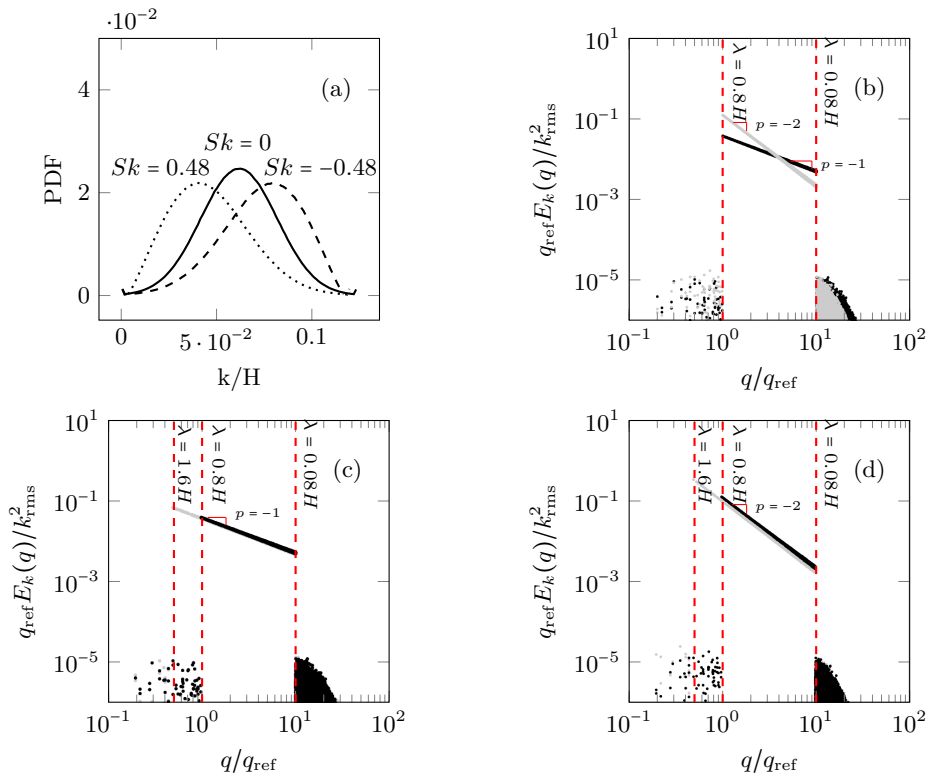

    \centering
    \begin{subfigure}[t]{.49\linewidth}
    \centering
     \input{tikz/HPDG14F.tikz}
    \end{subfigure}\hfill%
    \begin{subfigure}[t]{.49\linewidth}
        \centering
    \input{tikz/PSforp}
    \end{subfigure}
    \begin{subfigure}[t]{.49\linewidth}
        \centering
    \input{tikz/PSforl}
    \end{subfigure}\hfill%
    \begin{subfigure}[t]{.49\linewidth}
        \centering
    \input{tikz/PSforl2}
    \end{subfigure}
    \caption{Statistical representation of the studied roughness. (a): PDF of roughness, (b): Normalized PS density with different $p$, (c): Normalized PS density with different $\lambda_0$, $p=-1$, (d): Normalized PS density with different $\lambda_0$, $p=-2$. In (b,c,d) wavenumber $q$ is normalized by the reference wavenumber $q_{\text{ref}}=2\pi/(0.8H)$. Vertical dashed lines are high-pass filtering and low-pass filtering, corresponding to $\lambda_0$ \& $\lambda_1$ respectively. }
    \label{fig:HPDS}
\end{figure}
Using the roughness generation algorithm introduced in section \ref{sec:gen}, multiple samples are generated.
In the present work, different types of PDF will be combined with power-law PS, that is $E_k(\textbf{q})=C_0(\norm{\textbf{q}}/q_0)^p$, where \textbf{q} is the wavenumber vector, $\textbf{q}=(q_x,q_z)^\intercal$, $q_0=2\pi/\lambda_0$ is the smallest wavenumber corresponding to the largest in-plane length scale $\lambda_0$, $C_0$ is a constant to scale the roughness height, and $p$ is the spectral slope of the power-law PS.
An overview of the configurations of PDF and PS is illustrated in figure~\ref{fig:HPDS}.
In figure~\ref{fig:HPDS} (b,c,d) the PS density normalized by the root mean square (r.m.s.) of the roughness height are compared in pairs.
The upper and lower cutoff wavelengths $\lambda_0$ and $\lambda_1$ are transformed to cutoff wavenumbers $q_0=2\pi/\lambda_0$ and $q_1=2\pi/\lambda_1$, which are represented by the red dashed lines in figure~\ref{fig:HPDS} (b,c,d) on the left and right side of the figures respectively.
As stated in the previous section, the lower cutoff wavelength is related to the grid resolution and a value of $\lambda_1=0.08H\approx8\Delta_{x}\approx8\Delta_{z}$ is applied for all roughness topographies in the present work. With an isotropic roughness and a fixed $\lambda_1$, the PS is determined by two remaining parameters, $\lambda_0$ and $p$.
In present work, two values of $p$ ($p=-1$ and $p=-2$) are examined, the PS of which are shown in figure~\ref{fig:HPDS} (b).
For the selection of $p$ values we seek similarity to previous works~\citep{anderson_meneveau_2011,BARROS20181,nikora2019}.
Moreover, two different upper cutoff wavelengths ($\lambda_0=0.8H$ and $\lambda_0=1.6H$) of the roughness PS are investigated. 
Power spectrum with $\lambda_0=0.8H$ and $1.6H$ with identical slopes $p$ are compared in figure~\ref{fig:HPDS}(c,d), where wavenumber $q$ is normalized by referencing wavenumber $q_{\text{ref}}=2\pi/(0.8H)$

Three types of PDFs with positive, zero and negative skewness are examined in the present work.
Covering a relatively large range of $Sk$ is intended to ensure that the results can be generalized to a wide spectrum of naturally occurring roughness in different applications.
The non-skewed roughness is described by a Gaussian distribution.
For the positively skewed roughness, Weibull distribution is used, which can be written as
\begin{equation}
f(k)= K \beta^K k^{(K-1)}e^{-(\beta k)^K}~,
\end{equation}
\begin{figure}
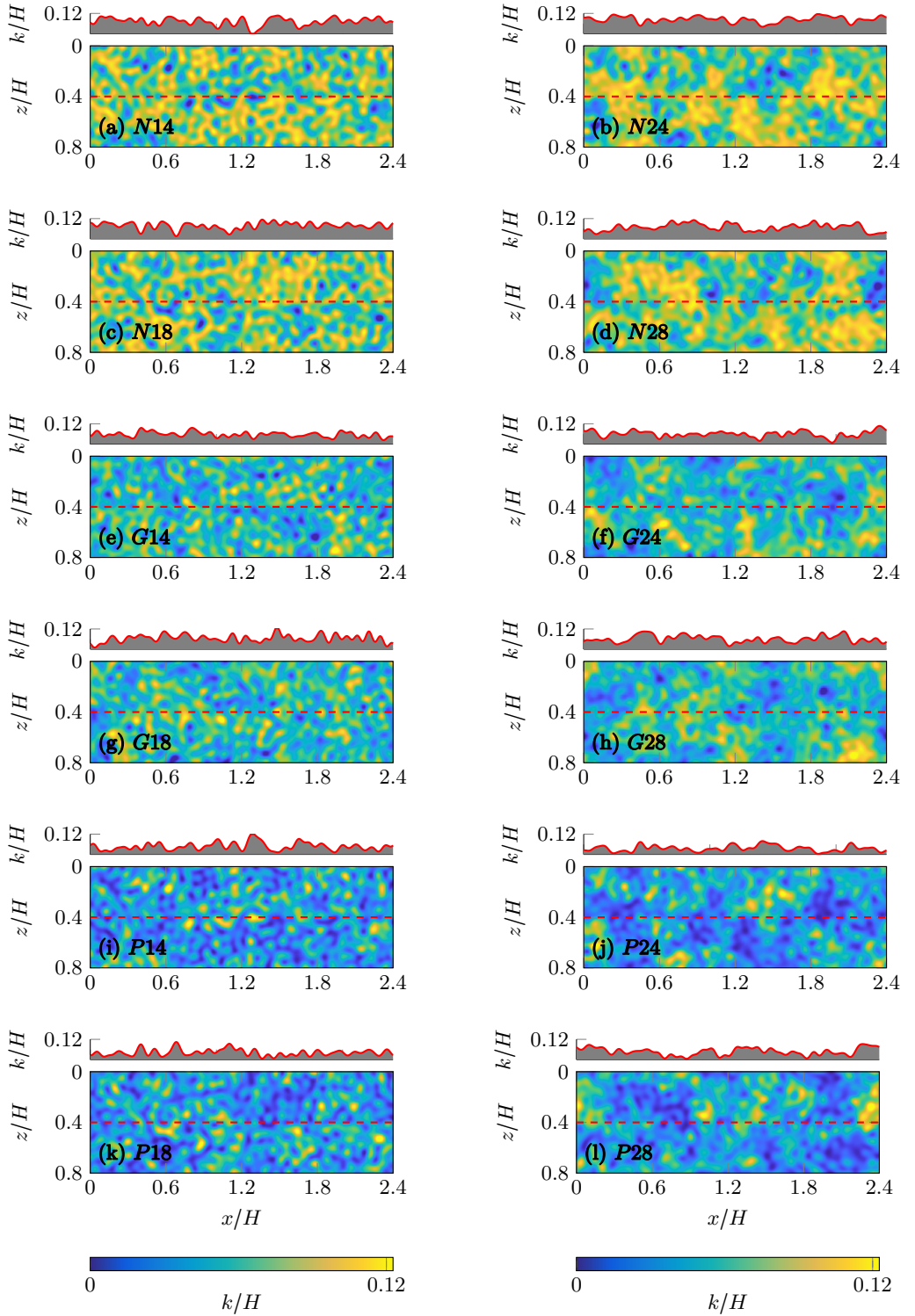

\centering
\begin{subfigure}{.45\linewidth}
\input{tikz/N14.tikz}
\end{subfigure}
\hfill
\begin{subfigure}{.45\linewidth}

\input{tikz/N24.tikz}
\end{subfigure}
\begin{subfigure}{.45\linewidth}
\input{tikz/N18.tikz}
\end{subfigure}
\hfill
\begin{subfigure}{.45\linewidth}

\input{tikz/N28.tikz}
\end{subfigure}

\begin{subfigure}{.45\linewidth}
\input{tikz/G14.tikz}
\end{subfigure}
\hfill
\begin{subfigure}{.45\linewidth}

\input{tikz/G24.tikz}

\end{subfigure}
\begin{subfigure}{.45\linewidth}
\input{tikz/G18.tikz}

\end{subfigure}
\hfill
\begin{subfigure}{.45\linewidth}
\input{tikz/G28.tikz}
\end{subfigure}
\begin{subfigure}{.45\linewidth}
\input{tikz/P14.tikz}
\end{subfigure}
\hfill
\begin{subfigure}{.45\linewidth}
\input{tikz/P24.tikz}
\end{subfigure}
\begin{subfigure}{.45\linewidth}
\input{tikz/P18.tikz}
\end{subfigure}
\hfill
\begin{subfigure}{.45\linewidth}
\input{tikz/P28.tikz}
\end{subfigure}
\vspace{-0.5cm}
\caption{Roughness maps with configuration $M1-500$, color indicates height. 1D roughness profile at $z=0.4H$ (\Rdashed{}) is shown above each roughness map. (a-d): negatively skewed, (e-h): zero skewness, (i-l): positively skewed; left column: $p=-1$. right column, $p=-2$.}
\label{AllPM1}
\end{figure}
where the parameters $K$ and $\beta$ can be used to adjust the standard deviation and skewness of the distribution. The skewness is always adjusted to the value of 0.48. Similar to the Gaussian distribution the kurtosis of the Weibull distribution is always equal to three.
A negatively skewed PDF is obtained by flipping the PDF of a positively skewed Weibull PDF. Here $Sk=-0.48$ is prescribed. In the present work, the $99\%$ confidence interval of roughness height PDF, $k_{99}$, is used as the characteristic size of roughness, i.e. $k=k_{99}$. This measure is related to the standard deviation of the roughness, and hence can be directly prescribed. We used a fixed value of $k=0.1H$ in all cases. Indeed, $k$ is a statistical measure of maximum peak-to-trough roughness size, which unlike the absolute peak-to-trough size $k_\text{t}$, is not deteriorated by extreme events.
These types of PDFs are illustrated in figure~\ref{fig:HPDS}.
Moreover, in order to avoid extreme high roughness elevations in the simulations, roughness heights outside 1.2 times the $99\%$ confidence interval of PDF are excluded.
Therefore, the peak-to-trough height $k_\text{t}=0.12H$ is achieved for all roughness in the present work.
Combining the three types of PDF (with different values of $Sk$) with four types of PS (two values of $p$ and $\lambda_0$ each) twelve different roughness topographies are studied in the present work, which are summarized in table \ref{tab:SumOfCase}.
Selected patches of all 12 roughness topographies 
on surfaces with size $2.4H \times 0.8H$, are displayed in figure~\ref{AllPM1}.
Above each roughness map, the 1D roughness profile at $z=0.4H$ along streamwise direction is shown.

For each roughness sample, simulations in full-span and minimal channels are carried out. For minimal channels the spanwise size $L_z$ is the main subject of the study. 
As the log-layer flow structures are set by the spanwise dimension $L_z$~\citep{Flores2010}, it is often most critical in terms of reducing the computational cost.
The spanwise size of the present minimal channels are designed to fulfil the three criteria set by inequalities (\ref{MINI2}). The first criterion ($L_z^+>100$) stems from the fact that the computational domain must accommodate the near-wall cycle of turbulence and, unlike the other two criteria, is independent of the roughness topography. The second criterion ($L_z^+\geq\tilde{k}^+/0.4$) ensures that roughness can be included in the 'healthy turbulent' zone under the critical height.
We translate the criteria for a realistic roughness by replacing $\tilde{k}$ for the sinusoidal roughness by the characteristic roughness height $k$ for any arbitrary roughness. 
The third criterion states that the minimal channel should contain the roughness wavelength.
However, for realistic surfaces a single characteristic wavelength is not naturally determined.
A conservative choice for this limit of the channel width can be the largest in-plane length scale, which is $\lambda_0$ in the present study. This ensures that all wavelengths present in the roughness topography are included in the spanwise domain.
Recalling the aim of reducing the cost of the roughness simulation, however, we seek a less conservative choice, in which some of the larger wavelengths are excluded.
Particularly, for simulations of engineering roughness, it is often impracticable to include extremely large roughness scales. 
To formalize our choice, we denote the largest spanwise wavelength that a domain can accommodate as $\lambda^*$, and calculate the portion of surface energy that larger wavelengths contribute to the original roughness as
\begin{equation}
\Phi_c\left(\frac{2\pi}{\lambda^*}\right)=\frac{\int_{2\pi/\lambda^*}^{2\pi/\lambda_1}E_k(q)\mathrm{d}q}{\int_{2\pi/\lambda_0}^{2\pi/\lambda_1}E_k(q)\mathrm{d}q}~.
\end{equation}
where $\lambda$ is the discrete wavelength. 
If the spanwise domain size is $\lambda^*$, the simulation resolves a roughness with $\Phi_c$ portion of the original surface variance.

In the current research we examine a choice of spanwise channel size corresponding to half the size of the largest length scale, i.e. $\lambda_0/2$. With the adopted power-law PS, it leads to the values of $\Phi_c$ equal or larger than 90\% for all samples under investigation.
Hence the third criterion is replaced by $L_z\geq\lambda_{1/2}$, where $\lambda_{1/2}=\lambda_0/2$, which means that the simulations resolve at least 90\% of the original surface variance.
The new criterion leads to a minimum channel width of $L_z=0.8H$ for half of the investigated roughness topographies (those with $\lambda_0=1.6H$) and $L_z=0.4H$ for the other half (those with $\lambda_0=0.8H$).
We label the minimal channels with the former (larger) and latter (smaller) spanwise sizes as $M$1 and $M$2, respectively.
For roughness samples with $\lambda_0=0.8H$ simulations at both $M$1 and $M$2 channels are carried out.
To complete the investigation, some simulations with a further reduced channel size $M$3 ($L_z=0.3H$) are carried out. 
It is worth mentioning that, since $k=0.1H$ holds for all topographies, $M$1, $M2$ and $M$3 channels fulfill the $L_z\geq k/0.4$ criterion. For all simulation configurations the streamwise channel size $L_x$ is set according to the equation \ref{MINI2} once $L_z$ is known.

Simulations are carried out at four different friction Reynolds numbers Re$_\tau=250$, $500$, $750$ and $1000$, at fixed $k/H=0.1$, leading to $k^+=25$, $50$, $75$, and $100$. However the parametric study on roughness topography is only conducted at $k^+=50$. Apart from minimal channel simulations, conventional full span channel simulations with the size $L_x\times L_y\times L_z=8H\times2H\times4H$, labeled as $F$, are also carried out for all roughness topographies. For the two highest Reynolds numbers, however, such large simulations are costly. Consequently, for these Reynolds numbers, the largest investigated channels are smaller than the $F$ channel (but still larger than $M$1/$M$2/$M$3). These channels are labeled as $M$0. Table \ref{tab:Re} summarizes the details of all simulations carried out for rough channels. In order to provide a reference for determining the roughness function $\Delta U^+$, additional smooth-wall simulations are also performed in $M2$, $M1$ and $F$-sized channels at Re$_\tau=500$ (not shown in the table). Overall, each rough-wall simulation case is defined by a combination of roughness topography and simulation configuration (channel size and Reynolds number). Throughout the article, the following naming convention is used to describe the cases: 
\begin{equation}
    \overbrace{\underbrace{\boxed{\;G\;}}_\text{PDF}\;\;\underbrace{\boxed{ \;2\;}}_\text{$-p$} \underbrace{\boxed{\;4\;}}_\text{\;$10\lambda_{1/2}/H$}}^{\text{Topography}}\;|\overbrace{\underbrace{\boxed{\;F\;}}_{\text{ch. size}}- \underbrace{\boxed{500}}_{10k^+}}^{\text{Simulation configuration}}~.
\end{equation}
\begin{itemize}
    \item The first character indicates the type of PDF; $G$ for Gaussian distribution, $P$ for positively skewed ($Sk\approx0.48$), and $N$ for negatively skewed ($Sk\approx-0.48$).
    \item The second digit indicates the PS spectral slope; $1$ for $p=-1$ and $2$ for $p=-2$.
    \item The third digit represents the half of large cutoff wavelength $\lambda_{1/2}$, with which channel width is determined; $4$ for $\lambda_{1/2}=0.4H$ and $8$ for $\lambda_{1/2}=0.8H$.
    \item The following character(s) indicates the channel spanwise size; $F$ for full channel ($L_z=4H$), $M$1 and $M$2 for the larger ($L_z=0.8H$) and the smaller ($L_z=0.4H$) minimal channels, respectively. $M$3 utilizes the spanwise width $L_z=0.3H$.
    $M$0 is introduced to the cases which $k^+=75$ and $100$ with $L_z=2H$ and $1H$, respectively
    \item The last number denotes $10k^+$, or equivalently, Re$_\tau$.
\end{itemize}

    \begin{table}
        \centering
        \begin{tabular}{cccccccccc}
            Topography & $Sk$ & $p$ & $\lambda_0/H$&$k_\text{a}/H$&$k_{\text{md}}/H$&$k_{\text{rms}}/H$&ES&$\Delta U^+$&$d/k$\\[3pt]
            $P14$ & $0.48$ & $-1$ & 0.8&0.017&0.046&0.0208&0.57&7.33&0.81\\
            $P18$ & $0.48$ & $-1$ & 1.6&0.017&0.046&0.0208&0.54&7.23&0.81\\
            $P24$ & $0.48$ & $-2 $& 0.8&0.017&0.046&0.0208&0.44&6.99&0.78\\
            $P28$ & $0.48$ & $-2 $& 1.6&0.017&0.046&0.0208&0.39&6.57&0.76\\[3pt]
            $G14$ & $0$ & $-1$ & 0.8&0.016&0.061&0.0200&0.54&6.67&0.95\\
            $G18$ & $0$ &$ -1$ & 1.6&0.016&0.061&0.0200&0.53&6.56&0.95\\
            $G24$ & $0$ & $-2$ & 0.8&0.016&0.061&0.0200&0.43&6.30&0.92\\
            $G28$ & $0$ & $-2$ & 1.6&0.016&0.061&0.0200&0.37&5.94&0.90\\[3pt]
            $N14$ & $-0.48$ & $-1$ & 0.8&0.017&0.074&0.0208&0.57&6.14&1.06\\
            $N18$ & $-0.48$ & $-1$ & 1.6&0.017&0.074&0.0208&0.54&5.84&1.06\\
            $N24$ & $-0.48$ & $-2$ & 0.8&0.017&0.074&0.0208&0.44&6.09&1.03\\
            $N28$ & $-0.48$ & $-2$ & 1.6&0.017&0.074&0.0208&0.39&5.51&1.01\\
        \end{tabular}
                \caption{Roughness topographical properties; skewness
        $Sk=(1/k_{\text{rms}}^3)\int_S(k-k_{\text{md}})^3\mathrm{d}S$, effective slope $\text{ES}=(1/S)\int_S|\partial k/\partial x|\mathrm{d}S$, mean absolute height $k_\text{a}=(1/S)\int_S|k-k_{\text{md}}|\mathrm{d}S$, root mean square height $k_{\text{rms}}=\sqrt{(1/S)\int_S(k-k_{\text{md}})^2\mathrm{d}S}$. Melt-down height $k_{\text{md}}=(1/S)\int_Sk\mathrm{d}S$ is measured from deepest trough and $S$ is the wall-projected surface area. The values of $\Delta U^+$ and $d/k$ are computed in full channels at Re$_\tau=500$.
        }
        \label{tab:SumOfCase}
    \end{table}

\begin{table}
    \centering

    \begin{tabular}{ccccccccccc}
         Topographies&Configuration&Re$_\tau$&$L_x/H$&$L_z/H$&$N_x$&$N_z$&$\Delta_x^+$&$\Delta_z^+$&$\Delta_{y,k}^+$ & FTT \\[3pt]
         $G24$&$M2-250$&250&4.0&0.4&512&48&1.95&2.08&0.88&1200\\
         $G24$&$M1-250$&250&5.0&0.8&576&96&2.17&2.08&0.88&300\\
         $G24$&$F-250$&250&8.0&4.0&900&480&2.22&2.08&0.88&100\\[3pt]
         $G24$&$M3-500$&500&2.0&0.3&256&48&3.91&3.13&1.74&1000\\
         $**4$ \& $G28$&$M2-500$&500&2.0&0.4&256&48&3.91&4.17&1.74&2000\\
         all&$M1-500$&500&2.4&0.8&256&96&4.69&4.17&1.74&500\\
         all&$F-500$&500&8.0&4.0&900&480&4.44&4.17&1.74&80\\[3pt]
         $G24$&$M2-750$&750&1.4&0.4&288&96&3.65&3.13&2.59&1200\\
         $G24$&$M1-750$&750&2.4&0.8&480&160&3.75&3.75&2.59&300\\
         $G24$&$M0-750$&750&4.0&2.0&640&320&4.69&4.69&2.59&100\\[3pt]
         $G24$&$M2-1000$&1000&1.2&0.4&288&96&4.17&4.17&3.53&500\\
         $G24$&$M1-1000$&1000&2.4&0.8&576&192&4.17&4.17&3.53&300\\
         $G24$&$M0-1000$&1000&3.0&1.0&720&240&4.17&4.17&3.53&100\\
    \end{tabular}
            \caption{Summary of all simulation cases including roughness topography and simulation configurations. For all cases $L_y/H=2$,$N_y=401$. Moreover, $**4$ indicates all roughness topographies with $\lambda_{1/2}=0.4H$, $\Delta_{y,k}^+$ indicates the grid size at the roughness height i.e. $y=0.1H$, flow through time (FTT$=TU_b/L_x$) for statistics collection duration are shown in the last column, where $T$ is the total integral time, $U_b=(1/H_{\text{eff}})\int_0^{H}U\mathrm{d}y$ is the bulk velocity, $H_{\text{eff}}=H-k_{\text{md}}$.}
            \label{tab:Re}
\end{table}

\subsection{Post-processing}
\label{sec:post}
The time-averaged flow field over a rough surface is heterogeneous in horizontal directions.
In order to analyze the one-dimensional (1-D) mean profile of the flow, we apply double averaging, as proposed by~\citet{SHAW198251}.
The double-averaged velocity profile in the wall-normal direction $\bigl<\overline{u}\bigr>(y)$ is obtained by averaging the time-averaged velocity over wall-parallel directions, i.e.
\begin{equation}
    \bigl<\overline{u}\bigr>(y)=\frac{1}{S}\iint_S\overline{u}(x,y,z)\mathrm{d}x\mathrm{d}z~.
\end{equation}
where $\overline{u}(x,y,z)$ is time averaged streamwise velocity, $S$ is the wall-normal projected plan area (i.e. area of the corresponding smooth wall) and angular bracket $\bigl<\cdot\bigr>$ denotes horizontal averaging.
The double-averaged velocity profile $\bigl<\overline{u}\bigr>(y)$ will be denoted as $U$ for simplicity.
The time-averaged velocity field is obtained over a long enough period of time.
It is reported by~\citet{Flores2010} in their study of smooth minimal channel that, due to the bursting events, simulation time required to achieve converged flow statistics is longer than conventional full span simulation . In order to achieve converged mean velocity profile, a minimum flow-through-time (FTT) is chosen to be 300 for minimal channels. The mean velocity profiles are proven converged for all the cases with 300 FTTs. 
Initial transients are removed from the statistical integration.

The influence of roughness on the mean flow can be accounted for by a modified coefficient of viscosity $\nu_e$ beyond the region where the shape of velocity profile is affected by roughness, i.e. outer layer \citep{perry_joubert_1963}. This $\nu_e$ can be interpreted as a downward shift in logarithmic layer in the inner-scaled streamwise velocity profile relative to the smooth case. This downward velocity shift in the logarithmic region is referred to as roughness function $\Delta U^+$~\citep{CLAUSER19561,hama1954}, which is further confirmed by a number of roughness studies, e.g.~\citep{Schultz2009}.
In other words, as a result of the outer layer similarity~\citep{Townsend}, which states that outer-layer flow is unaffected by the near wall events except for the effect due to the wall shear stress, the downward shift of the velocity profile is approximately a constant value in the logarithmic region and possibly beyond if the outer-flow geometry and Reynolds number are matched.
Introducing the roughness function to the logarithmic law of the wall (log-law hereafter), it writes
\begin{equation}
        U^+=\frac{1}{\kappa}\text{ln}(y^+-d^+)+B-\Delta U^+~.
\end{equation}
where $\kappa\approx0.4$ is the von Kármán constant, $\textit{B}\approx5.2$ is the log-law intercept for the smooth wall, $d$ indicates the zero plane displacement which will be talked in detail in the following section, and the superscript $+$ indicates scaling in wall units.
Based on the pioneering work by~\citet{Nikuradse1933}, the roughness function in the fully rough regime is a sole function of the inner-scaled roughness size $k_s^+$ for the sand-grain roughness according to
\begin{equation}
    \Delta U^+ =B-8.48+\frac{1}{\kappa}\text{ln}(k_s^+)~.
    \label{asysptot}
\end{equation}
Equation~\ref{asysptot} is the basis for the definition of `equivalent' sand-grain roughness (also denoted by $k_s$) for an arbitrary roughness with the same roughness function.

In the present work, the roughness function $\Delta U^+$ is calculated as the mean offset of the inner-scaled mean velocity profile over the logarithmic layer for the cases with Re$_\tau\approx500$.
Since corresponding smooth channels in $M2$, $M1$ and $F$ with matched Re$_\tau\approx500$ are available, this quantity is calculated by direct comparison to the smooth case.
For those cases with varied Re$_\tau$,  further smooth channel simulations with matched Re$_\tau$ are required in order to derive the corresponding profile shift, which causes unfavourable computational effort. 
Having in mind that log-law applies for minimal smooth channels under critical height $y_c$, a good approximation of velocity profile in log region of the smooth channels can be drawn from the log-law, thus $\Delta U^+$ is estimated by the velocity shift at the critical height $y_c^+=0.4\times L_z^+$ relative to the log-law $U^+=(1/\kappa)\text{ln}(y^+)+5.2$, where $\kappa=0.4$.

Finally it must be noted that, unlike a smooth channel, the origin of the wall-normal coordinate for the log-law is not naturally defined for a rough wall.
In this regard, \citet{jackson_1981} suggested use of moment centroid of the drag profile on rough surface as the virtual origin for the logarithmic velocity profile.
The definition of the virtual wall zero-plane displacement $d$ in present work follows Jackson's method.

\section{Results}
\label{DNS}
\subsection{Evaluation of minimal-channel simulations}
\label{sec:EMC}
As reviewed in the previous section, while a few of the previous studies applied the minimal channel approach for irregular roughness, they did not systematically examine it for this type of roughness. For instance in the work by~\citet{jouybari_2021}, a validation case is carried out by duplicating the irregular roughness structure in both wall-parallel directions. With such an approach, the minimal channel effect considers only the effect of fluid domain expansion or reduction but might lack the roughness-related effect due to the repetitiveness of the rough surface.
In order to comprehensively assess the applicability and limits of the minimal channel, first the simulations of the roughness topography $G24$ with variation of channel size at matched $k^+\approx50$ is discussed in \cref{width}. This is followed by the results for all different roughness topographies in \cref{minimal channel-roughness topography}.
As mentioned before, the roughness generation process in the present research has a random nature, where only statistical properties are prescribed. To understand if a mere `statistical representation' can lead to a unique flow response, simulations are carried out for eight random realizations of the roughness topography $G24$ with $k^+\approx50$. In studying different random realizations with identical PDF and PS we believe we could have contributed to answering this fundamental question. The results are presented in~\cref{Rando}.
Finally in \cref{Res}, one roughness topography is studied in a wide range of roughness Reynolds numbers ($k^+=25-100$) in order to assess the prediction of the minimal channel in different rough regimes. Here the roughness topography $G24$ is evaluated using minimal channels $M$2 and $M$1 and (pseudo) full-span channels $F/M$0. 
\subsubsection{Minimal channels with different spanwise sizes}
\label{width}
\begin{figure}
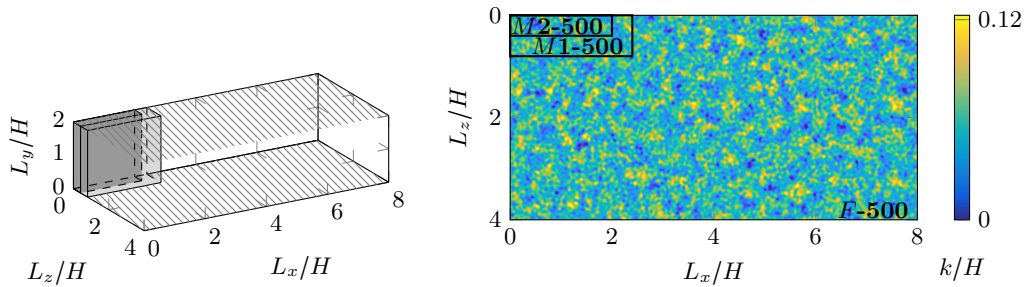

\centering
\begin{subfigure}[t]{.42\linewidth}
\input{tikz/3Droughness}
\end{subfigure}\hfill
\begin{subfigure}[t]{.57\linewidth}
\input{tikz/SurfaceG24F.tikz}
\end{subfigure}
\caption {Comparison of channel sizes. Left: Schematic simulation domain of $F-500$, $M1-500$ and $M2-500$, hatch pattern represents roughness. Right: Roughness map of $G24F-500$, black frames indicate minimal channels $M1-500$ \& $M2-500$}
\label{G24F}
\end{figure}

\begin{figure}
    \centering
    \input{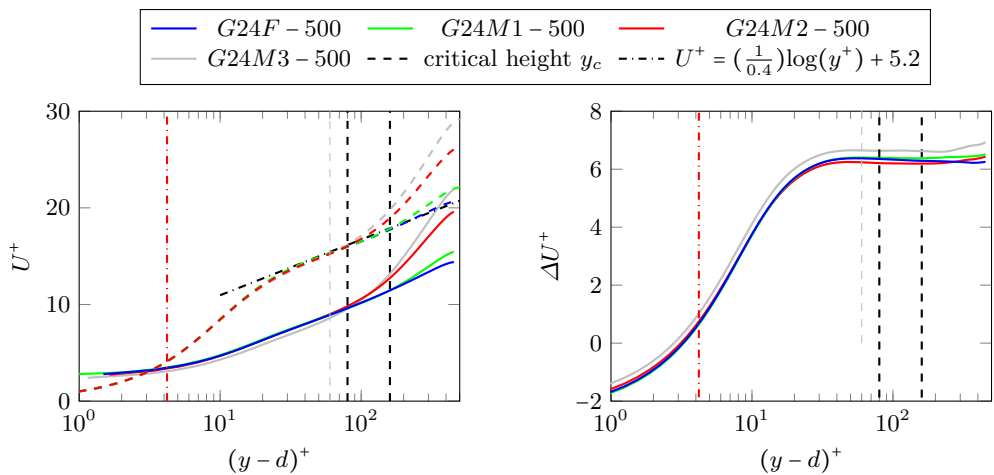}
    \caption{Simulation results of roughness type $G24$. Left: mean velocity profiles (\full: Rough, \dashed: Smooth), right: velocity offset profiles. Red vertical line indicates roughness height measures from the zero-plane displacement $(k-d)^+$.}
    \label{UG24}
\end{figure}
The 3D schematic representations of minimal channels $M2$ and $M1$ as well as the full span channel $F$ used for simulations at $k^+\approx 50$ are shown in figure~\ref{G24F} (left); $M3$ is not shown for simplicity.
The hatched pattern indicates where the roughness is mounted.
Roughness topography $G24$ in the full-size simulation is shown in figure~\ref{G24F} (right). 
For a direct comparison, boundaries of minimal channels $M1$ and $M2$ are represented by the black frames.
The pseudo-random surfaces for each configuration is generated independently.
That is, for a specific topography, the surface height map in each simulation is unique, but they all share identical statistical properties.

The inner-scaled velocity profiles obtained from roughness topography configuration $G24$ with $k^+\approx50$ are shown in figure~\ref{UG24} (left). 
Colored dashed lines are the velocity profiles extracted from smooth channel. 
The color indicates the size of channel.
The critical heights of minimal channels $M2-500$ and $M1-500$, i.e. $y_c^+=0.4\times L_z^+=80\text{ and }160$ are illustrated by black vertical dashed lines in the figure, respectively.
While the critical height of $M3-500$ i.e. $y_c^+=60$ is shown with gray vertical dashed line. It can be observed from the figure, that minimal channel cases $M2-500$ and $M1-500$ successfully reproduce the velocity profile of a conventional full-span channel under the critical height $y_c$. 
The velocity profiles deviate above the critical height $y_c$ due to the nature of minimal channels.
For $G24M3-500$, however, some discrepancy of the profile can be observed even under its critical height.
In figure~\ref{UG24} (right), the velocity offset profiles for $F$, $M1$ and $M2$ channels are displayed. The velocity offset profiles are obtained by subtracting the rough channel velocity profile from each corresponding smooth channel velocity profiles.
The velocity offset profile from minimal channels $M1$, $M2$ and full-span channel on the right panel show excellent agreement.
Consequently, it seems like the velocity offset is not meaningfully affected by absence of the large wavelengths in spanwise direction with small contribution to the roughness height power spectrum.
This however does not hold for channel $M3$ where $\Phi_c(2\pi/L_z)=82\%$. A similar parameter study performed for roughness topography $G28$ (not shown here) revealed that the velocity offset starts to deviate for channel $M2$ ($\Phi_c(2\pi/L_z)=84\%$). In both cases the deviation of the velocity profiles starts when the contribution of excluded large wavelengths in the roughness height spectrum is larger than 10\%.
In contrast to the minimal domain guidelines Eqn.~\ref{MINI2} that were based on the single sinusoidal wavelength, the generalisation of this idea is not entirely clear for irregular roughness, which contains a wide range of wavelength. Following the findings that the large, undulating scales do not contribute to drag~\citep{BARROS20181}, the present paper demonstrates that an \textit{a priori} rule of thumb is 90\% of the surface variance. This is checked \textit{a posteriori} by comparing the coherence spectra between surface variance and drag to show that the drag-carrying physics are resolved which will be discussed in~\cref{IBMF}.

In the present work, $\Delta U^+$ of minimal channels are obtained by averaging the mean velocity offset from each critical height $y_c^+$ to the half of channel half-height $0.5H^+$, while full span channels is averaged in the region $y+=80-250$.
This gives $\Delta U^+=6.3$ for case $G24F-500$ and $\Delta U^+=6.4$ and $6.2$ for cases $G24M1-500$ and $G24M2-500$, respectively.

\subsubsection{Minimal channels for different roughness topographies}
\label{minimal channel-roughness topography}
\begin{figure}
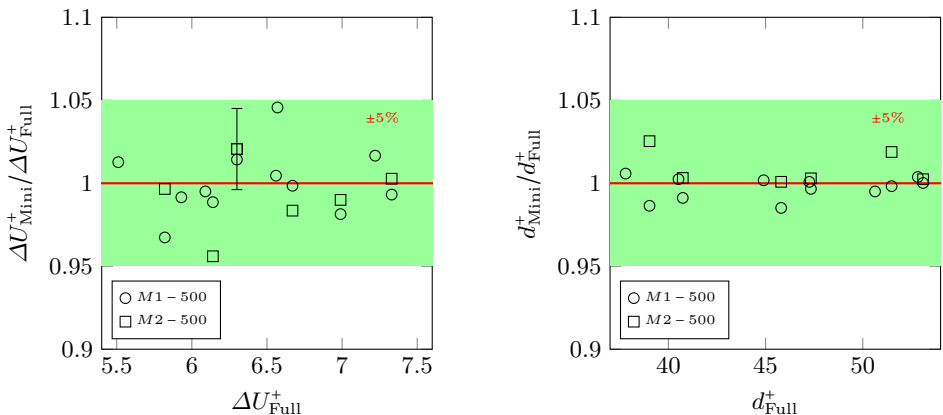

     \centering
     \begin{subfigure}[t]{.49\linewidth}
     \input{tikz/DUFullmini2}
     \label{DUfullmini}
      \end{subfigure}
      \begin{subfigure}[t]{0.49\linewidth}
      \input{tikz/dFullmini2}
       \end{subfigure}
      \caption{Roughness function (left) and zero plane displacement (right) predicted by minimal channels normalized with full span channel prediction, $\circ$: $M1-500$, $\square$: $M2-500$. Green shadow indicates prediction error interval of $\pm5\%$. In left panel, $99\%$ confidence interval for different simulations of case $G24M2-500$ is shown as an error bar.}
      \label{fullmini}
 \end{figure}

Applying the same analysis to all topographies at $k^+\approx50$, roughness function $\Delta U^+$ are calculated and represented in figure \ref{fullmini} (left). In this figure, $\Delta U^+$ predicted by minimal channels are compared with the prediction by full-span channels with matched topographical property.
It has to be mentioned that for case $G24M2-500$, multiple simulations are carried out for the purpose that will be discussed in~\cref{Rando}.
Therefore, roughness function of $G24M2-500$ is the mean roughness function $\overline{\Delta U^+}$ over $G24M2-500$s. 
In figure \ref{fullmini} (left), the $\pm5\%$ disagreement interval is illustrated by the green shadow around $\Delta U_{\text{Mini}}^+/\Delta U_{\text{Full}}^+=1$ (red line).
Another key quantity widely discussed in the framework of roughness studies is the zero-plane displacement $d$. Similar to $\Delta U^+$, $d$ is often used as input to roughness models, and therefore, its prediction is of practical value. 
Predicted zero-plane displacements $d$ in minimal channels are compared with full span channels in figure~\ref{fullmini} (Right).
It can be observed that minimal channel predictions show excellent agreement with conventional full span channel, the discrepancy lies under 5\%.

Consistent predictions of roughness function $\Delta U^+$ indicate the capability of the minimal channels in reproducing roughness function $\Delta U^+$ of the irregular pseudo-realistic roughness even if a certain range of larger wavelengths are excluded. Obviously, in minimal channel simulations, large turbulent structures in the outer layer cannot be resolved; therefore an non-physical wake behaviour is observed in the outer layer of minimal channels mean velocity profile (see figure~\ref{UG24}). However, as the present results suggest, capturing the near wall turbulence in the minimal channel is adequate for the prediction of roughness function $\Delta U^+$ as a quantification of skin friction drag. This, obviously, cannot be generalized to all aspects of the turbulent flow.
    
 \subsubsection{Effect of randomness}
 \label{Rando}
 {In the present work, roughness is generated following a pseudo-random process with prescribed PDF and PS. As a result, individually generated rough surfaces with identical statistics are not deterministically identical.}
 This randomness can be a source of uncertainty when pseudo-random roughness is used as a surrogate of realistic roughness (and possibly explaining the scatter observed in figure~\ref{fullmini}).
 {The pseudo-random roughness generation process can also be considered an imitation of random roughness formation processes in the nature or industry. Hence it can be used to shed light on whether a statistical representation of stochastic roughness is adequate to predict the flow response.}
 
 To this end, eight rough surfaces corresponding to $G24$ topography are generated independently.
 Realization of each randomly generated surface is unique while the statistical properties are virtually identical. 
 The averaged value of roughness function over the eight samples calculated at $k^+\approx 50$ in minimal channel $M$2 is $\overline{\Delta U^+}_{G24M2-500}=6.43$, while the 99\% uncertainty interval of all values is 0.31.
 This averaged value is shown in figure~\ref{fullmini} along with the uncertainty interval.
 One can observe that the uncertainty bar well encompasses the $\Delta U^+$ in the full channel. {This can be taken as an indication that minimal channel prediction can approximately converge to the exact value if the main uncertainty due to randomness is ruled out. Nevertheless, as stated before, the error associated with one random realization is still considerably low. Additionally, one cannot rule out a minor influence due to other factors, e.g. the nature of turbulence in the minimal channel, but the present data suggest those influences to be minor if present.} It is observed in figure~\ref{fullmini} that the 99\% confidence bar lies in the green area -- the 5\% error range.
 Overall, it can be stated that DNS in minimal channels with matched roughness statistics is an accurate tool for the prediction of $\Delta U^+$ of realistic roughness, apart from the small discrepancy, which is arguably linked to the effect of randomness.\\
 {It is appropriate at this point to recall that PDF and PS are effectively `reduced order' representations of the actual roughness geometry. The current results suggest that this representation can almost uniquely determine the dynamic response of flow for the studied type of roughness topographies as far as $\Delta U^+$ is concerned. However, since a reduced model do not contain all information, minor differences in $\Delta U^+$ among surfaces with the same PDF and PS is not an unexpected observation. In other words, while the generated roughness is controlled in a global sense by PDF and PS, local distribution of its features can be affected by randomness. For example, occurrence of clustered or streamwise aligned roughness peaks can lead to attenuation of drag due to the sheltering effect, which will be discussed in section~\ref{IBMF}}.




\subsubsection{Minimal channels in transitionally and fully rough regimes}
\label{Res}

\begin{figure}
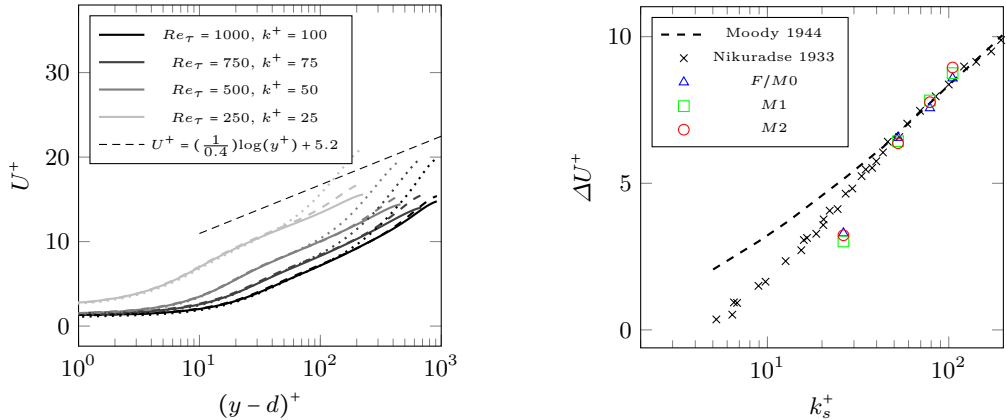

    \centering
    \begin{subfigure}[t]{0.45\linewidth}
    \centering
        \input{tikz/Re_expand}
    \end{subfigure}\hfill%
    \begin{subfigure}[t]{0.45\linewidth}
    \centering
        \input{tikz/KsRe}
        \vspace{0.4cm}
    \end{subfigure}
    \caption{Simulations at different $k^+$ at fixed $k/H$; $k^+=25-100$. Left: Mean velocity profiles, line color gradually changes from gray to black with increasing Re$_\tau$. (\full: $F/M0$, \dashed: $M1$, \dotted: $M2$). Right: Roughness function. Data from Nikuradse's uniform sand grain roughness and Colebrook relation provide for industrial pipes are added for comparison.} 
    \label{Re_expand}
\end{figure}
The values of $\Delta U^+$ reported for the simulations with $k^+\approx50$ suggest that the flow likely lies in the border between transitionally and fully rough regimes. In order to ensure that minimal channels deliver acceptable predictions in a wide range of scenarios including both regimes, in this section we study one roughness topography ($G$24) in a range of roughness Reynolds numbers $k^+\approx25-100$. 
Both minimal channel simulations $M2$ and $M1$ as well as large-span channel simulations $F/M0$ are carried out. 
Simulation setups are summarized in table~\ref{tab:Re}.
Mean velocity profiles are shown in figure~\ref{Re_expand} (left), while roughness functions $\Delta U^+$ against $k_s^+$ are shown in figure~\ref{Re_expand} (right). The inner-scaled equivalent roughness height $k_s^+$ on the abscissa of the latter figure is obtained by scaling the calculated $k_s$ (as explained below) with viscous length scale $\delta_\nu$ at different Re$_\tau$.
One can observe from figure~\ref{Re_expand} (left) that each velocity profile deviates above the respective critical height $y_c^+=0.4\times L_z^+$ which are not shown for clarity. 
Based on these velocity profiles, roughness functions are obtained by calculating the velocity difference at each critical height relative to the log-law $U^+=(1/0.4)\text{log}(y^+)+5.2~$.
Equivalent sand-grain roughness $k_s$ is calculated by fitting roughness function to the asymptotic roughness function in fully rough regime of Nikuradse sand-grain roughness as shown in figure~\ref{Re_expand} (right). In doing so, we obtain an equivalent sand-grain roughness size of $k_s\approx1.05k$ for roughness topography $G24$. Notably, the calculated values of roughness function from both minimal and full channels show an excellent agreement. 
Furthermore, it can be observed that $\Delta U^+$ asymptotically approaches fully rough regime at $\Delta U^+\approx6$ for both minimial and full channels.

\subsection{Roughness surface force}
\label{IBMF}

\begin{figure}
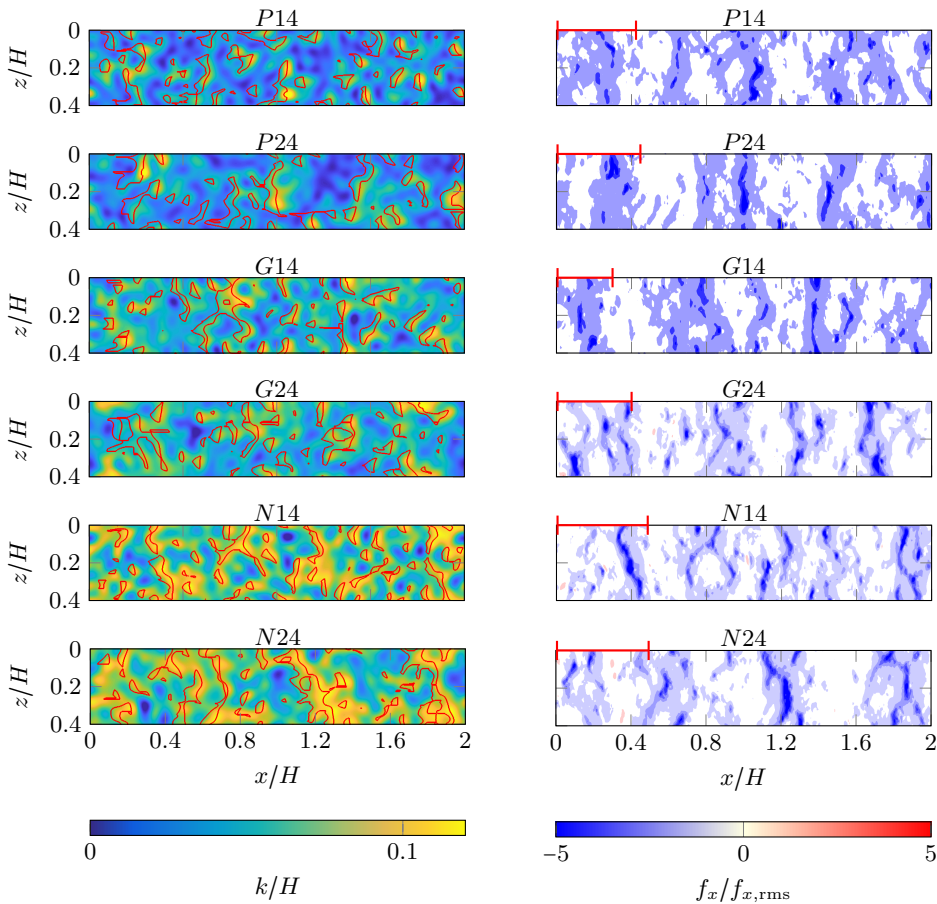

    \centering
    \begin{subfigure}{0.49\linewidth}
    \centering
    \input{tikz/P14_mask}
    \end{subfigure}
    \begin{subfigure}{0.49\linewidth}
    \centering
    \hspace{-2.4mm}
    \input{tikz/P14_force}
    \end{subfigure}
        \begin{subfigure}{0.49\linewidth}
    \centering
    \input{tikz/P24_mask}
    \end{subfigure}
    \begin{subfigure}{0.49\linewidth}
    \centering
    \hspace{-2.4mm}
    \input{tikz/P24_force}
    \end{subfigure}
        \begin{subfigure}{0.49\linewidth}
    \centering
    \input{tikz/G14_mask}
    \end{subfigure}
    \begin{subfigure}{0.49\linewidth}
    \centering
    \hspace{-2.4mm}
    \input{tikz/G14_force}
    \end{subfigure}
    \begin{subfigure}{0.49\linewidth}
    \centering
    \input{tikz/G24_mask}
    \end{subfigure}
    \begin{subfigure}{0.49\linewidth}
    \centering
    \hspace{-2.4mm}
    \input{tikz/G24_force}
    \end{subfigure}
            \begin{subfigure}{0.49\linewidth}
    \centering
    \input{tikz/N14_mask}
    \end{subfigure}
    \begin{subfigure}{0.49\linewidth}
    \centering
    \hspace{-2.4mm}
    \input{tikz/N14_force}
    \end{subfigure}
        \begin{subfigure}{0.49\linewidth}
    \centering
    \input{tikz/N24_mask}
    \hspace{2.8mm}
    \end{subfigure}
    \begin{subfigure}{0.49\linewidth}
    \centering
    \vspace{-3.1mm}
    \input{tikz/N24_force}
    \hspace{0.2mm}
    \end{subfigure}
    \caption{Roughness (left column) and surface force distribution (right column) pairs in $M$2. The exposed surface derived from the 1-D sheltering model is marked by red contour line on the roughness distribution maps. Separation lengths of force peaks obtained from the auto-correlation are represented by the red bars on the upper left corner of the surface force maps.}
    \label{G24Force_Dist}
\end{figure}

In the present simulations, IBM introduces a volume force within the solid area imposing zero velocity and hence represents the action of pressure and viscous drag force.
One of the advantages of IBM is the explicit representation of localized hydrodynamic force exerted by roughness~\citep{chan-braun_garcia-villalba_uhlmann_2011}, here referred to as `surface force'.
In this section, we investigate the link between the mean surface force distribution and the roughness height distribution.
Given the satisfactory performance of the minimal channel demonstrated in~\cref{sec:EMC}, the following analysis is carried out based on the results achieved from minimal channels.
Previous studies on irregular roughness report that a certain range of roughness scales is dominant in generation of skin friction~\citep{BARROS20181}.
A deeper insight into the contribution of different roughness scales to the drag force is the aim of this section. To this end, the local forcing map $\textbf{f}(x,z)$ is obtained by time-averaging the IBM force field $\textbf{f}_{\text{IBM}}(x,y,z,t)$ and integrating the force in wall-normal direction $y$:
\begin{equation}
    \textbf{f}(x,z)=\frac{1}{T} \int_0^H\int_0^T\textbf{f}_{\text{IBM}}(x,y,z,t) \mathrm{d} t \mathrm{d} y,
\end{equation}
{\textbf{f}$(x,z)=(f_x,f_y,f_z)^\intercal$ is the force vector and $f_x$, $f_y$ and $f_z$ are streamwise, wall-normal and spanwise force component, respectively.} { One should note that, precisely speaking, \textbf{f} equals force per unit density and wall-projected area. Nevertheless, as we are interested in its trend rather than its absolute value, this quantity always appears in a normalized form; hence, for brevity we refer to it as `force'.}
The visualization of normalized forcing map for all $M2-500$ cases with their roughness distribution maps are shown in figure~\ref{G24Force_Dist}.
The entire set of surface force distributions show spanwise-elongated coherent areas of negative forcing. 

\begin{figure}
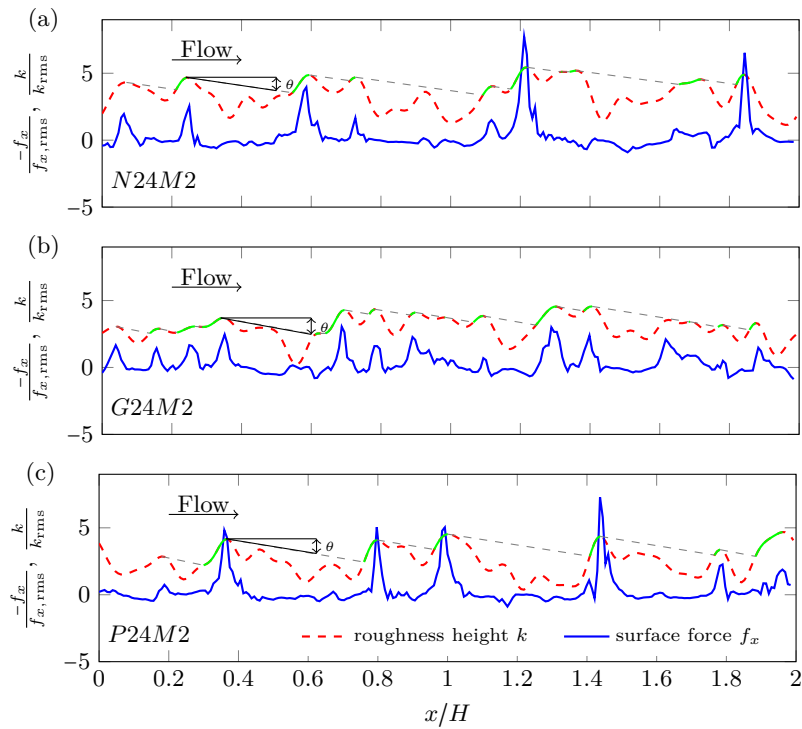

    \centering
        \begin{subfigure}{\linewidth}
           \centering
        \input{tikz/N24distVSforce}
    \end{subfigure}
    \begin{subfigure}{\linewidth}
        \centering
        \input{tikz/G24distVSforce}
    \end{subfigure}
    \begin{subfigure}{\linewidth}
        \centering
        \input{tikz/P24distVSforce}
    \end{subfigure}
    \caption{Normalized force $f_x$ and roughness distribution profile for $N$24$M$2 (upper), $G$24$M$2 (middle) and $P$24$M$2 (lower) at $z=0.2H$. Sheltering effect modeled by \citet{yang16} is illustrated by gray dashed lines, the unsheltered surfaces are marked by green profile.}
\label{distforce}
\end{figure}
\begin{figure}
  \centering
    \input{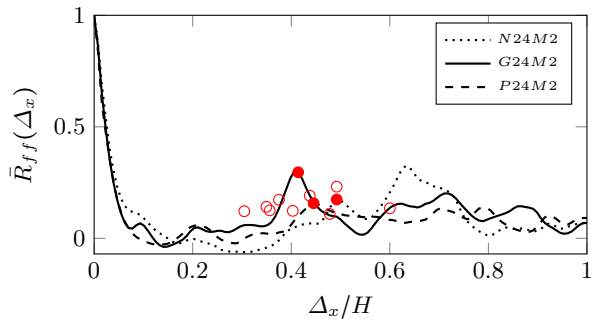}
  \caption{Auto-correlation function of the streamwise surface force component as a function of streamwise separation $\Delta_x$. The first correlation peaks of the three auto-correlation functions are marked by red points, the peaks of the rest studied cases are marked by red circles.}
    \label{Autocorrs}
\end{figure}
Furthermore, force distributions in streamwise direction at $z=0.2H$ are displayed for three cases $N$24$M$2$-$500 (negative skewness), $G$24$M$2$-$500 (Gaussian) and $P$24$M$2$-$500 (positive skewness) in figure~\ref{distforce} along with the surface height functions at the same location.
In this figure, solid blue line represents the normalized negative force profile, ${-f_x}/{f_{x,\text{rms}}}$, while dashed red line represents the corresponding normalized roughness profile, ${k}/{k_{\text{rms}}}$.
Comparing different roughness topographies, it can be observed that the Gaussian surface demonstrates a larger number of extreme force peaks than the surfaces with asymmetric PDF.
As expected, the peaks in surface force are mostly collocated with the peaks in roughness height. A sudden rise in the negative force is expected when the mean flow impinges on the windward side of the roughness element followed by a rapid drop on the leeward side, which is also observed in the figure. Interestingly, the force peaks are much narrower than the surface height peaks, which can be attributed to the separation of flow behind the roughness peak. Another notable observation is that the pronounced force peaks show a much longer streamwise separation than the height peaks.
Such observation can be linked to the sheltering effect, which causes a significant reduction in the flow momentum in the wake of a tall roughness element. 

\citet{yang16} investigated the sheltering effect on the surface roughened by rectangular-prism roughness elements and argued that once the region sheltered by the upstream roughness element covers the neighbouring elements, the surface drag decreases.
They suggested that an attenuation parameter for skin friction should incorporate the `shadowed area'.
To provide further insight into the present observations, we apply the wake expansion model proposed by these authors -- with some simplification -- to the roughness profiles in figure~\ref{distforce}.
According to~\citet{yang16}, the streamwise slope of the sheltered region from separation point down to the ground can be calculated from the wake expansion rate with $\tan\theta=C_\theta u_\tau/U_h$, where $u_\tau$ is the friction velocity, $U_h$ is the velocity at roughness element height,
$C_\theta=1-(2/3)(1-h/w)$ is the shape parameter of the roughness and $h/w$ denotes the aspect ratio of the rectangular prisms.
 {With the aim to investigate the underlying physical mechanism of the sheltering model, we expand the use of the model -- which is obtained based on rectangular prisms roughness -- to more realistic irregular roughness.}
To approximate the expansion rate of irregular roughness in the present work we use double-averaged mean velocity at each roughness peak height as $U_h$. We also replace $h$ with the characteristic roughness height $k_{99}=0.1H$ and $w$ with the spanwise integral length scale of the roughness $L_{k,z}\approx 0.05H$, which will be defined in the following section.
Using these values, $C_\theta=1.7$ is obtained and the resulting shadowed area in figure~\ref{distforce} is indicated by the gray dashed lines. In the same figure, the `exposed' areas on the roughness peaks are highlighted by green lines.
It is clear that the extreme force peaks coincide almost exclusively with the green areas and the shadowed areas rarely produce any significant local force. This can be an indication on the applicability of the wake expansion model to irregular roughness.
The 1-D sheltering model is applied to the 2-D roughness distribution in figure~\ref{G24Force_Dist}.
    The exposed roughness surface are outlined by the red contour lines.
    One can observe that the exposed surface contours match well with the localization of the surface force.
    {Notably, the spanwise elongated patterns of the surface force and their streamwise separation can be well reproduced with the help of sheltering model. This finding can also be an indication of the feasibility to predict the local drag with a knowledge of roughness structure, which has a predictive potential for more complex problems, e.g. inhomogenous and anisotropic roughness structures.}

To shed further light on the surface force patterns and sheltering effect, the streamwise auto-correlation functions of the streamwise surface force $\bar{R}_{ff}(\Delta_x)$ for $N24M2$, $G24M2$ and $P24M2$ cases are shown in figure~\ref{Autocorrs}.
The streamwise auto-correlation function of surface force is defined as
\begin{equation}
\bar{R}_{ff}(\Delta_x)=\frac{1}{L_xL_z}\int_{0}^{L_z}\int_{0} ^{L_x} \frac{f_x(x,z)}{f_{x,\text{rms}}} \frac{f_x(x+\Delta_x,z)}{f_{x,\text{rms}}}\mathrm{d}x\mathrm{d}z~.
\end{equation}
A rapid drop of auto-correlation in the vicinity of zero separation -- a result of the narrow peaks in the surface force distribution -- is observed in figure~\ref{Autocorrs}.
Additionally, a mild but clear positive peak in auto-correlation function, as marked by red circles, is observed at a  separation of approximately 0.3-0.6$H$. This value is likely to be related to the streamwise distance between the force peaks.  
Locations of the second auto-correlation peaks for the rest of studied cases are marked by hollow red circles in the same figure without showing the auto-correlation functions for better clarity. 
For visual comparison, we also indicate these values by red bars on the upper left corner of the respective surface force maps in figure~\ref{G24Force_Dist}. Here it can be confirmed that, roughly speaking, the length of the bars are similar to the separation between the spanwise-elongated areas with high surface force.

\subsubsection{Correlation between surface force and roughness height}
\label{correlation}
In this section, the link between streamwise component of surface force, $f_x(x,z)$, and roughness height distribution, $k(x,z)$ is analysed by means of correlation function of the two quantities. 
The correlation function $R_{kf}(\Delta_{x})$ is calculated along the streamwise direction followed by averaging in the spanwise direction:
       \begin{equation}
    \bar{R}_{kf}(\Delta_x)=\frac{1}{L_xL_z}\int_{0}^{L_z}\int_{0} ^{L_x} \frac{\left(k(x,z)-k_{\text{md}}\right)}{k_{\text{rms}}}\frac{\left(f_x(x+\Delta_x,z)-\bar{f}_{x}\right)}{f_{x,\text{rms}}}\mathrm{d}x\mathrm{d}z,
    \label{eq:Rkf}
\end{equation} 

where the subscript rms and overbar indicate root mean square value and mean value, respectively.
The calculated values of correlation coefficient $\bar{R}_{kf}(\Delta_x=0)$ for the studied topographies are summarized in Table~\ref{Cro-Coef}. The negative sign of correlation indicates that high roughness peaks are correlated with negative surface force (force directed against the streamwise mean flow), as expected.
It is observed that the negatively skewed topographies show lower correlation coefficient, which can be linked to the fact that this type of roughness is rather prone to generation of recirculation and separation zones in the surface valleys and indentations, so the responding force is less localized in those areas. 
Oppositely, for the positively skewed topographies higher correlation coefficients are observed due to the peak-dominated structures, in which the protruding parts of roughness are directly responsible for generation of localized drag force.
The correlation coefficients of the exposed surface with the surface force distribution, $\bar{R}_{kf,\text{exp}}$, are also estimated and shown in table~\ref{Cro-Coef}.
Hereby only the exposed surfaces, i.e. the surface area that are marked by red contour lines in figure~\ref{G24Force_Dist}, are kept on the height map, while the sheltered surfaces are replaced by 0 elevation.
Noticeable increase in the correlation coefficient can be observed for all cases, especially for negatively skewed roughness, where the correlation is increased by approximately 35\%, while the increase for Gaussian and positively skewed roughness are 20\% and 16\%, respectively.
This, once again, highlights the importance of sheltering effect and exposed roughness areas for the generation of surface force.

\begin{table}
    \centering

    \begin{tabular}{ccc|ccc}
         Case&$\bar{R}_{kf}$&$\bar{R}_{kf,\text{exp}}$&Case&$\bar{R}_{kf}$&$\bar{R}_{kf,\text{exp}}$  \\[3pt]
         $P14M2-500$&-0.46&-0.51&$P18M1-500$&-0.48&-0.54\\
         $P24M2-500$&-0.48&-0.56&$P28M1-500$&-0.47&-0.56\\ [3pt]
         $G14M2-500$&-0.44&-0.56&$G18M1-500$&-0.46&-0.56\\
         $G24M2-500$&-0.44&-0.51&$G28M1-500$&-0.45&-0.59\\ [3pt]
         $N14M2-500$&-0.38&-0.54&$N18M1-500$&-0.42&-0.55\\
         $N24M2-500$&-0.41&-0.55&$N28M1-500$&-0.43&-0.57\\
    \end{tabular}
                \caption{Cross-correlation coefficient $\bar{R}_{kf}(\Delta=0)$ and $\bar{R}_{kf,\text{exp}}(\Delta=0)$}
    \label{Cro-Coef}
\end{table}

\begin{figure}
    \centering
    \input{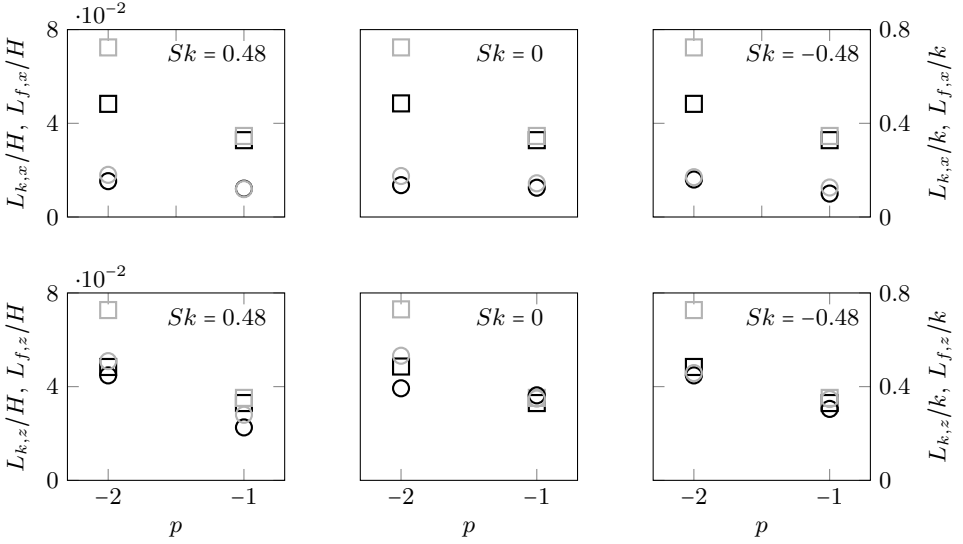}
    \caption{Integral length scale $L_k$ and $L_f$ as functions of $p$, grouped by $Sk$. Left axis: normalized by $H$, right axis: normalized by $k$. The squares represent $L_k$ while circles represent the $L_f$. Black: $\lambda_0=0.8H$, gray: $\lambda_0=1.6H$. Upper row: streamwise integral length, lower row: spanwise integral length.}
    \label{IntLength}
\end{figure}
Furthermore, we calculate the integral length scales of streamwise surface force ($L_f$) and roughness height distribution ($L_k$). 
The integral length scales are calculated in a similar way as proposed by~\citet{Quadrio2003} using the following expression for the integral length scale of roughness height
\begin{equation}
L_{k}=\int_{\Delta=0}^{L^{\text{corr}}_{0.2}}\bar{R}_{kk}(\Delta) \mathrm{d}\Delta ,
\end{equation}
where $\Delta$ is the separation in either streamwise or spanwise directions and $L^{\text{corr}}_{0.2}$ is the separation at which the auto-correlation function drops under the arbitrary value of $0.2$.
The integral length scales for surface force $L_f$ is computed in a similar fashion. The calculated values can be regarded as a scale for the width of roughness elements or force peaks. Both integral length scales are calculated for different cases and plotted in figure~\ref{IntLength} as functions of topographical properties, i.e. PS slope $p$ and $\lambda_0$.
In these figures, streamwise integral length scales are plotted on the upper row, while spanwise integral length scales are plotted on the lower row, grouped by $Sk$.
Square symbols represent $L_k$ while circles represent $L_f$.
It is observed that surface force has a smaller streamwise integral length scale than the surface height, which is in line with the qualitative observation of the very narrow force peaks in figure~\ref{distforce}.
As expected, topographical parameters show a clear impact on the integral length scales of roughness height.
Contrary to this, the streamwise integral length scale of force $L_{f,x}$ does not show strong sensitivity to the considered roughness variation.
A notable observation is that the spanwise integral length scale of force $L_{f,z}$ is more sensitive to the topographical changes than the streamwise length scale. The value of $L_{f,z}$ is comparable to the surface height integral length scale $L_{k,z}$. 
{Based on the limited data points in our dataset, the two quantities show similar trends with $p$ and (to some extent) $\lambda_0$: a higher value of $L_{f,z}$ is obtained for the roughness with $p=-2$ and $\lambda_0=1.6H$.}
Unlike the isotropic behavior of roughness height function, clearly illustrated by the comparable streamwise and spanwise integral length scales, the distribution of surface force is observed to be strongly anisotropic.
The fact that the integral length scale of surface force is different in $x$- and $z$-directions is the quantitative manifestation of spanwise-elongated coherent areas of surface force observed in figure~\ref{G24Force_Dist}.

\subsubsection{Coherence function of surface force and roughness}
\label{coherence}

\begin{figure}
    \centering
    \input{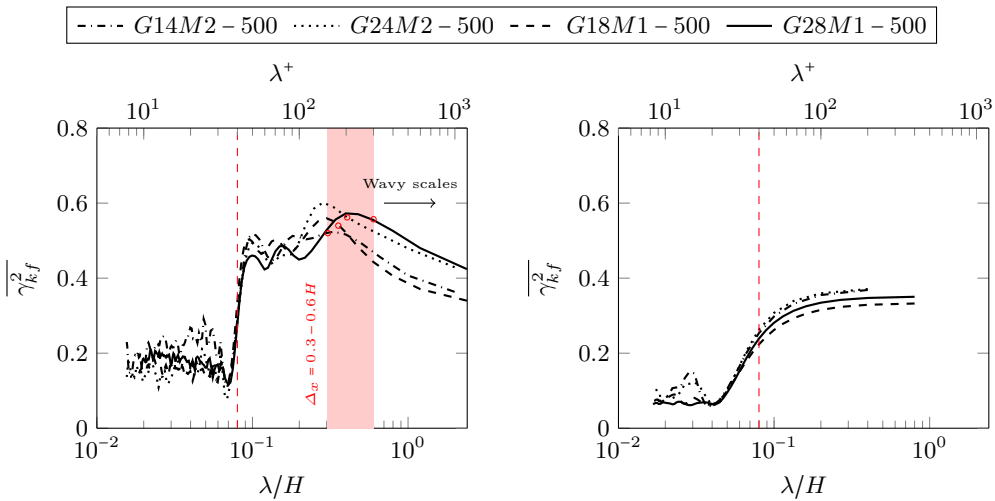}
    \caption{Mean coherence function $\bar{\gamma}_{kf}^2$ as a function of $\lambda=2\pi/q$ normalized by $H$. Left: streamwise, right: spanwise. $\lambda_1=0.08H$ is marked by vertical dashed lines on the left and right side, respectively. $\lambda^+$ for the current cases Re$_\tau\approx500$ is shown on the upper axis. The length scale of force separation obtained from its auto-correlation functions are marked by red circles on each coherence function respectively.}
    \label{Coh_Lambda}
\end{figure}

To further understand the correlation between force and height distributions at different roughness length scales, coherence function between the two distributions is calculated.
The coherence function represents the correlation of force distribution with roughness height distribution as a function of wavenumber:
\begin{equation}
    \gamma_{kf}^2(q)=\frac{|E_{kf}(q)|^2}{E_{f}(q)E_{k}(q)},
\end{equation}
where $E_{kf}(q)$ represents cross-PS of roughness topography $k(x,z)$ and force map $f_x(x,z)$, while $E_{f}(q)$ represents the PS of $f_x(x,z)$.
Power spectra are calculated based on 1D distribution profiles along streamwise and spanwise directions, and the mean coherence function $\overline{\gamma_{kf}^2}$ is obtained by averaging each parallel signal pairs.
Figure~\ref{Coh_Lambda} shows the mean coherence function of Gaussian surfaces in (a) streamwise direction and (b) spanwise direction as a function of wavelength $\lambda={2\pi}/{q}$ -- upper axis shows inner scaled wavelength $\lambda^+$ at Re$_\tau=500$.
It is worth reminding that for all topographies, the smallest in-plane roughness scale is prescribed to be $\lambda_1=0.08H$.
This is clearly related to the observation that coherence functions are significantly smaller below $\lambda\approx0.08H$ ($\lambda^+\approx40$).
Above this threshold, coherence functions increase and retain high values until a certain wavelength, which is roughly at $\lambda\approx0.3-0.6H$ ($\lambda^+\approx150-300$) for the studied cases.
With further evolution of the coherence function to larger wavelengths, the coherence decreases monotonically.
Similar observations are made for negatively and positively skewed roughness.
The force becomes less correlated with the surface features at very large scale, or in other words, very large wavelengths in streamwise direction do not contribute to the generation of surface force.
These length scales might be related to the `wavy roughness' concept stemming from the observations by~\citet{Schultz2009}. \citet{BARROS20181} stated that these length scales can be filtered out in regard of determining the skin friction.

\begin{figure}
    \centering
    \input{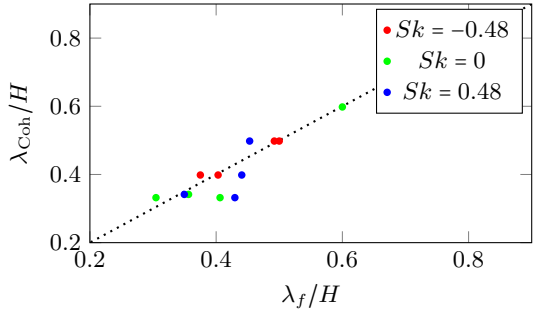}
    \caption{The length scale of force peak separation detected from the auto-correlation functions $\lambda_{f}$ compared with the coherence dropping wavelength $\lambda_{\text{Coh}}$.}
    \label{corrcohlength}
\end{figure}

Notably, the streamwise wavelength at which the coherence starts to drop has a similar value to the streamwise separation distance of the surface force peaks.
A comparison between the coherence dropping wavelength $\lambda_{\text{Coh}}$ and the length scale of force peak separation $\lambda_{f}$ is conducted in figure~\ref{corrcohlength}.
It can be observed that all data points are clustering around $\lambda_{\text{Coh}}=\lambda_{f}$ (dotted line) indicating a clear correspondence of these length scales.
The significant coherence at relative small length scales might be linked to the interaction of roughness structure and sheltering.
As discussed before, the occurrence of extreme force peaks is strongly determined by the roughness areas that are exposed in the high-momentum flow, or outside the sheltering.
Thus, streamwise recurring force peaks caused by sheltering can be found in figure~\ref{G24Force_Dist}.
Less prominent force peaks can be found between two successive extreme peaks, which contributes to the coherence function at small wavelength.
Furthermore, the roughness structures whose length scales are comparable to the distance between two successive extreme peaks, i.e. $\lambda_f$, show significance in the coherence function at corresponding wavelength.
Beyond this length scale, no larger force peak separation can be found, the coherence function keeps decreasing into large wavelength region.
Figure \ref{Coh_Lambda}(b) demonstrates that, unlike the tortuous behavior exhibited by the streamwise mean coherence function, the spanwise mean coherence function shows a monotonically increasing trend before a plateau at larger wavelengths. 
\subsection{Effect of roughness topographical properties}
\label{results}
In~\cref{sec:EMC}, we discussed applicability of minimal channel concept for characterization of realistic roughness. In doing so, we examined a relatively large number of roughness topographies, which provide a basis for studying the effect of roughness topography on hydrodynamic properties of the surface, which is discussed in the present section. It is already shown that PDF and PS can be considered as a reduced representation of roughness topography (almost) uniquely reproducing the hydrodynamic response. The common practice in the literature is, however, to parameterise roughness in terms of a few statistical parameters -- an even further reduced representation. Aiming to establish a link with the existing literature, we adopt and examine this approach in section 3.3. We first show the trends of $\Delta U^+$ and zero-plane displacement with some statistical parameters in section 3.3.1. Then some existing roughness correlations are assessed based on the present results in section 3.3.2.
 \subsubsection{Effect of statistical parameters on $\Delta U^+$ and $d$}
 
\begin{figure}
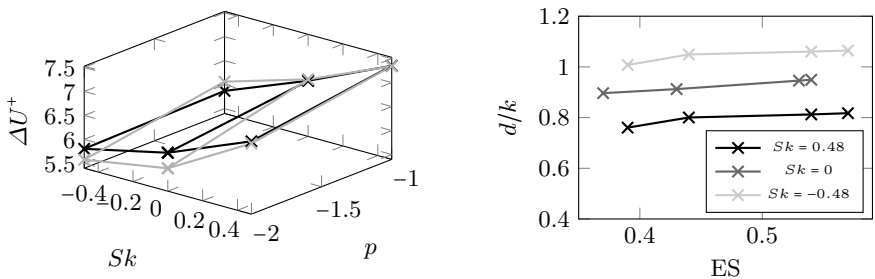

\centering
\begin{subfigure}{.45\linewidth}
\centering
    \input{tikz/plot3forl.tikz}
\end{subfigure}
    \begin{subfigure}{.45\linewidth}
    \centering
    \input{tikz/plot2dk}
\end{subfigure}
    \caption{Effect of roughness topographical properties to the $\Delta U^+$ prediction (left) from minimal channels, black: $\lambda_0=0.8H$, gray: $\lambda_0=1.6H$; $d$ predictions (right) as the function of ES, grouped by $Sk$.}
    \label{fig-3d}
\end{figure}

An overview of the roughness function $\Delta U^+$ for all topographies (see \cref{tab:SumOfCase}) is plotted in figure~\ref{fig-3d} (left), where roughness function $\Delta U^+$ is shown as a function of two of the investigated roughness parameters, i.e. skewness $Sk$ and PS slope $p$.
As investigated by~\citet{Flack2020}, positively skewed rough surfaces give higher skin friction than non-skewed or negatively skewed roughness.
{In general, $\Delta U^+$ reaches a higher value with $p=-1$.}
As illustrated in figure~\ref{fig:HPDS}(b), at $p=-2$ larger wavelengths contribute more to the roughness and at $p=-1$ \textit{vice versa}. 
These findings agree with the study by~\citet{BARROS20181} and the results in \cref{IBMF} highlighting that larger horizontal length scales contribute less to the hydrodynamic drag. It should be noted that the effective slope of roughness is larger for $p=-1$ compared to corresponding cases with $p=-2$.
Furthermore, it can be observed that the surfaces with $\lambda_0=1.6H$ show stronger sensitivity to the change of PS slope $p$ than those with $\lambda_0=0.8H$.

{Although the observed trends with selected statistical parameters can, to some extent, justify use of these parameters in predictive correlations, it is also observed that predicting skin friction based on a few roughness statistics is incomplete. Considering the roughness statistics and the values of roughness function in table~\ref{tab:SumOfCase}, it is observed that different types of roughness with similar statistical properties, e.g. the roughness with same values of $Sk$ and $k_{rms}$ in table~\ref{tab:SumOfCase}, show meaningful variation of their $\Delta U^+$ values. A better correlation for $\Delta U^+$ is observed when $ES$ is added to the prediction. However, inclusion of $ES$ is still not necessarily expected to yield unique predictions. A simple illustrative example could be that roughness formed by staggered and aligned roughness elements with identical statistics can lead to a significantly different values of in $k_s$, as shown in the study by~\citet{10.1115/1.4037280}.}

  As stated before, for a rough surface, the logarithmic layer of the flow is shifted upwards with respect to the bottom plane. Thus, the origin of wall-normal coordinate cannot be defined \textit{a priori}.
 As a result it is necessary to use a physically justified virtual origin for the logarithmic law of the wall. 
 The virtual origin lies above the $y=0$ plane at a distance equal to zero-plane displacement $d$.
 The value of zero-plane displacement $d/k$ following Jackson's method~\citep{jackson_1981} are documented in table~\ref{tab:SumOfCase}.
 To summarize the effect of roughness topography on zero-plane displacement, $d/k$ is plotted as a function of  effective slope ES on the right panel of figure~\ref{fig-3d}, while the data points are grouped by $Sk$.
 Even though ES is not explicitly prescribed in the present work, it is indirectly controlled by the two PS parameters $p$ and $\lambda_0$. 
 It can be observed that the value of $d/k$ increases with an increase in ES and a decrease in $Sk$, while the skewness effect is more dominant.

\subsubsection{Assessment of existing roughness correlations}
\label{sec:corr}
In this section, results from previously introduced topographies at $Re_\tau\approx500$ are used to assess some of the existing roughness correlations.
 As a matter of fact, existing roughness correlations are developed based on a limited number of data points covering a certain region of the parameter space~\citep{Chung2021annrev}. In this section, we are particularly interested to shed light on the generalization of these correlations outside their original parameter space, which is a key for a correlation to work across a wide range of rough surfaces encountered in different applications.

\begin{figure}
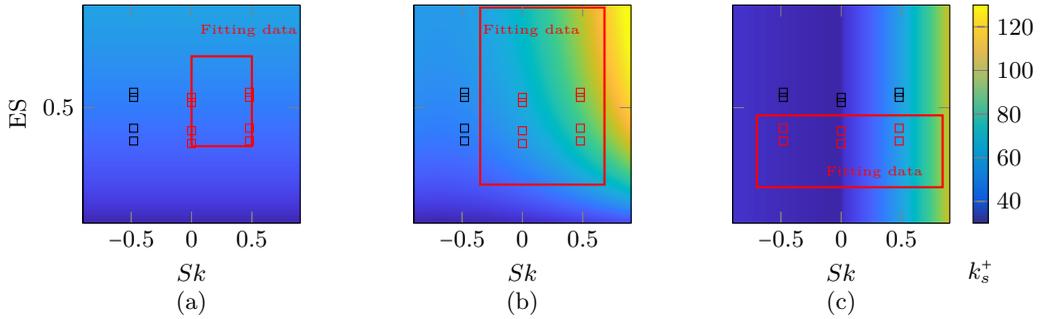

    \centering
        \begin{subfigure}[t]{0.3\linewidth}
        \input{tikz/CorrChan_Ks}
    \end{subfigure}\hfill%
    \centering
        \begin{subfigure}[t]{0.3\linewidth}
        \hspace{0.4cm}
        \input{tikz/CorrForooghi_Ks}
    \end{subfigure}\hfill%
    \begin{subfigure}[t]{0.3\linewidth}
        \input{tikz/CorrFlack_Ks}
    \end{subfigure}
    \caption{Predictive correlations for $k_s^+$. Squares indicate data points from present work while red frame represents the fitting data from literature.  (a): Eqn.~\ref{Chan2} by~\citet{chan2015}, (b): Eqn.~\ref{forooghi} by~\citet{10.1115/1.4037280}, (c): Eqn.~\ref{flack} by~\citet{Flack2020}.}
    \label{Cor2}
\end{figure}

In the following we assess 
three relatively recent correlations by~\citet{chan2015}, \citet{10.1115/1.4037280}, and ~\citet{Flack2020}, each predicting $k_s$ based on a few roughness statistical parameters. These correlations are applicable in the fully rough regime. Figure~\ref{Cor2} visualizes the selected correlations, where parameter space covered by the original fitting data of each correlation is represented by a red frame. In each sub-figure the data points from the present work are depicted as symbols. Different symbol colors are used to make distinction between the data points lying inside and outside the parameter space originally used for development of the correlation. Here the parameter space is expressed in terms of the two widely used parameters $Sk$ and ES


Required roughness statistics of the present roughness topographies are listed in Table~\ref{tab:SumOfCase}.
First we examine the correlation proposed by~\citet{chan2015} 
 \begin{equation}
     k_s=7.3k_a\text{ES}^{0.45}~.
     \label{Chan2}
\end{equation}
which is developed based on 3D sinusoidal roughness. In figure~\ref{Cor2} (a) the data points from the present work locate in the range of fitting data except for the topographies with $Sk<0$.
Obviously, since $Sk$ is not used as a predictive parameter in this correlation, it returns same predictions for different values of $Sk$.

Furthermore, two correlations developed by~\citet{10.1115/1.4037280}
 \begin{equation}
     k_s=k[0.67Sk^2+0.93Sk+1.3][1.07(1-e^{-3.5\text{ES}})]~,
     \label{forooghi}
 \end{equation}
and by~\citet{Flack2020}
 \begin{equation}
        \left\{
\begin{array}{ll}
      k_s=2.73k_{\text{rms}}(2+Sk)^{-0.45}~,& Sk<0~,\\
      k_s=2.11k_{\text{rms}}~,& Sk=0~,\\
      k_s=2.48k_{\text{rms}}(1+Sk)^{2.24}~,& Sk>0~, \\
\end{array} 
\right. 
\label{flack}
 \end{equation}
are examined. The original fitting data, from which these correlations are extracted, are also only partly cover the current data as illustrated in figure \ref{Cor2}.

\begin{figure}
    \centering
    \input{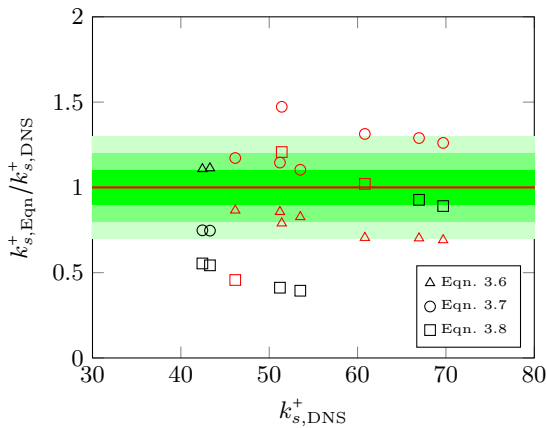}
    \caption{Prediction of $k_s^+$ compared to DNS. Red symbols indicate the present data points located within the fitting dataset. 10\%, 20\% and 30\% error intervals are represented by green shadow.} 
    \label{Corks}
\end{figure}

In order to directly evaluate the models, $k_s^+$ predicted by correlations are normalized and plotted against the full-span DNS results in figure~\ref{Corks}. 
The equivalent sand grain sizes $k_s^+$ of roughness from the simulations are obtained by fitting roughness function to the fully-rough asymptote, i.e. Eqn.~\ref{asysptot}. One should recall that the present data points can cover both transitionally and fully rough regimes, while the correlations in question are to be examined for the latter regime. Based on the result in section~\ref{Res}, and similar to the approach adopted by \cite{jouybari_2021}, an approximate value of $\Delta U^+\approx6$ is regarded as the threshold of fully-rough regime and only the data points in the fully-rough regime are shown in~\ref{Corks}. It has to be mentioned that this value is an approximate criterion to estimate the roughness regime. Exact $\Delta U^+$ criterion for different types of roughness can only be achieved through comprehensive experiments. 
In figure~\ref{Corks} different error intervals in prediction of $k_s^+$ are illustrated by green shades.
In general, a similar range of error can be observed among prediction of all correlations in the figure where  a limited number of data points lie outside the 30\% $k_s$ error area.
The fact that none of the correlations are able to perfectly reproduce the effect of topography on $k_s$ - as already pointed out by other authors ~\citep{Flack2020} - can be acknowledged in figure~\ref{Corks}. Among all correlations, the ones by Chan \textit{et al.} and Flack \textit{et al.} incorporate less geometrical information by taking one parameter related to the topography each, namely the effective slope of roughness in the former and the skewness in the latter. Forooghi \textit{et al.} combined both approaches. 
While the latter correlation delivers better predictions for some data points, an obviously superior accuracy cannot be established. One notable observation from all correlations is that their prediction does not deteriorate particularly for the data outside their original fitting range. This can be an indication that either of the models can be used with a similar level of reliability in a wider parameter space that is originally designed for.
Obviously, this statement is unlikely to hold for extreme cases outside the scope of this paper.

{To summarize, this section highlights the need for a more general model, e.g. by taking advantage of advances in data-driven methods. For the future development of roughness models/correlations, it is suggested by the present work that the roughness statistics that contain both height distribution and horizontal scales of roughness need to be incorporated for generalizable predictions.}


\section{Conclusion}
\label{Conclusion}
DNS is carried out for turbulent flow over irregular roughness in plane channels with reduced stream- and spanwise extents -- referred to as minimal channels. 
Roughness topography is mathematically generated using the method proposed by \citet{PEREZRAFOLS2019591}, in which PDF and PS of roughness map can be prescribed with high precision. Simulations are run for 12 different roughness topographies at $k^+=50$ and for a selected topography at $k^+=25-100$ (spanning both transitionally and fully rough regimes). For all cases, solutions are produced in full channels and one or more minimal channels. It is systematically demonstrated that, the value of roughness function for an irregular roughness with random nature can be predicted within $\pm5\%$ error using DNS in a minimal channel. This can be achieved as long as the minimal channel dimensions follow a relaxed version of the criteria suggested for regular sinusoidal roughness by previous authors \citep{chung_chan_macdonald_hutchins_ooi_2015,macdonald_chung_hutchins_chan_ooi_garcia-mayoral_2017}. The relaxation concerns the condition that the channel should contain all horizontal scales of roughness. Current data suggest that accurate prediction can be achieved as long as the size of channel is large enough to accommodate more than 90\% of original roughness height spectral energy based on the area under 2D PS. This finding is particularly relevant in DNS-based characterization of realistic rough surfaces that may contain very large wavelengths with limited contribution to the root mean square roughness height.

To shed  more light on possible origins of the mild discrepancy between minimal and full channel results, for one topography, multiple rough surfaces are generated. Due to random nature of roughness generation process, these surfaces are deterministically different while statistically identical. Simulations are carried out for these surfaces with $k^+=50$ and slightly scattered values of roughness function are obtained. Notably, roughness function value for the full channel resides within 99\% uncertainty interval of these scattered predictions. The results indicate that, at fixed PDF and PS, randomness in roughness generation can lead to a small uncertainty, which is also likely the origin of the observed $\pm5\%$ discrepancy between predictions of minimal and full channels.
This can be an indication that one can consider a combination of PDF and PS as a reduced order representation of roughness topography leading to nearly unique dynamic flow responses.

In addition to global flow properties, local surface forces for different types of roughness are calculated and their correlations with respective roughness height functions are studied.
It is observed that not all roughness height peaks generate force peaks. Applying the sheltering model proposed by~\citet{yang16} with some assumptions, we are able to show that only `exposed' (unsheltered) roughness peaks generate prominent peaks in surface force.
Notably, the spanwise elongated patterns of the surface force and their streamwise separation can be well reproduced with the help of the sheltering model.
This can be taken as a clear indication of the relevance of sheltering effect in flow over irregular roughness -- e.g. for complex terrains.\\

To shed light on contribution of different roughness scales to global drag, we also studied the coherence function of roughness height and surface force power spectra as a function of sreamwise and spectral scales. In streamwise direction, it was observed that coherence starts dropping beyond a certain length. This observation can be interpreted as smaller contribution of very large roughness wavelengths to the drag force. 
These large roughness length scales might be related the the `wavy roughness' concept stemming from the previous studies \citep{Schultz2009}.

Our analysis of surface force reveals certain previously unattended facts about roughness induced drag, e.g. reduced coherence between friction and roughness height at large scales.
Notably, the wavelength at which the coherence starts dropping is shown to be related to the separation between the peaks of surface force, which is linked to the sheltering effect itself. Unlike the streamwise direction, the coherence function does not drop in spanwise direction for the cases studied in this paper.

As stated above, present results suggest that an accurate yet computationally economical framework for characterization of irregular, realistic rough surfaces is in hand. Such a framework, can for example, be a basis for generation of large databases required for future `data-driven' roughness correlations. While this can be considered an obvious future research direction, in the present paper, we used the results from the 12 simulated roughness topographies to study the dependence of roughness function as well as zero-plane displacement on some key roughness parameters. Notably, it was shown that normalized zero plane displacement $d/k$ is most sensitive to the skewness of roughness distribution (larger at smaller values of skewness), and it also mildly increases with effective slope.

Finally we assessed a number of widely cited roughness correlations in the literature (Eqns.~\ref{Chan2}-\ref{flack}). While some correlations show a certain level of success in reproducing the roughness function or equivalent sand-grain roughness compared to the DNS results (see figure \ref{Corks}), there is an obvious need for improvement. An interesting observation is that none of the assessed correlations show a dramatic loss of accuracy when used outside the parameter space of its original fitting data. However, even the most successful correlations, can only reproduce the DNS data within $\pm30\%$ accuracy. This can arguably be the ground for a paradigm shift in development of future roughness correlations. As mentioned before a data-driven approach, which can account for the stochastic nature of roughness and its interaction with near-wall turbulence may be a solution to this problem. Recently, this idea has received some attention \citep{jouybari_2021,Brereton21} and more work in this direction is called for.
\section*{Acknowledgements}
Jiasheng Yang and Pourya Forooghi gratefully acknowledge financial support from Friedrich und Elisabeth Boysen-Foundation (BOY-151). This work was performed on the supercomputer ForHLR Phase 2 and the storage facility LSDF funded by the Ministry of Science, Research and the Arts Baden-Württemberg and by the Federal Ministry of Education and Research. The data presented in the manuscript are openly available in the KITopen repository at doi: \href{https://doi.org/10.5445/IR/1000142136}{10.5445/IR/1000142136}.

\section*{Declaration of interests}
The authors report no conflict of interest.

 \section*{Appendix A. Visualization of grid resolution}
 { In the present work, IBM is employed to introduce the roughness into the simulation flow field. With the help of IBM the complex geometry of the roughness can be represented on simple Cartesian grid. However, in order to completely represent the roughness with Cartesian grid, sufficiently fine grid resolution is essential. The grid resolution in the present work is illustrated in figure~21, where one of the rough surfaces with the smallest correlation length $L_x^{\text{corr}}$, i.e. $G14M2-500$ is shown. This rough surface contains the finest roughness structure among all the cases.}
     \begin{figure}
         \centering
         \input{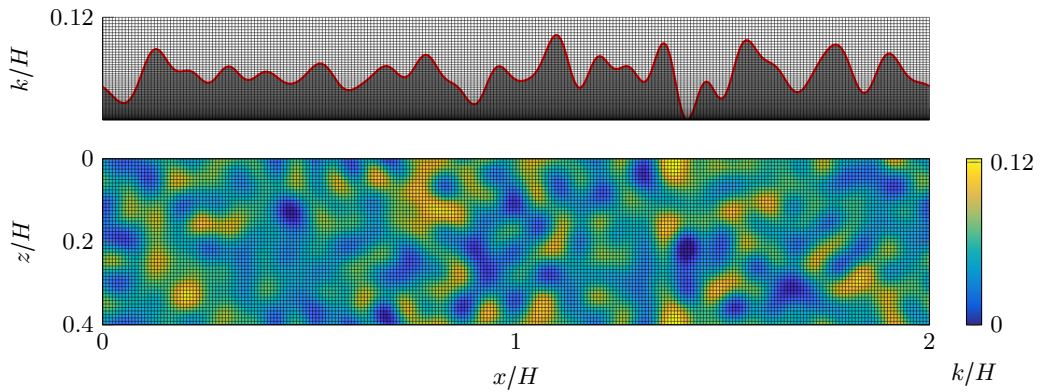}
         \caption{Mesh visualization in the wall layer with rough surface $G14M2-500$. Upper: mesh in wall-normal direction along $z=0.2H$. Lower: mesh in wall-parallel direction.}
         \label{Grid}
     \end{figure}
     \begin{figure}
         \centering
         \input{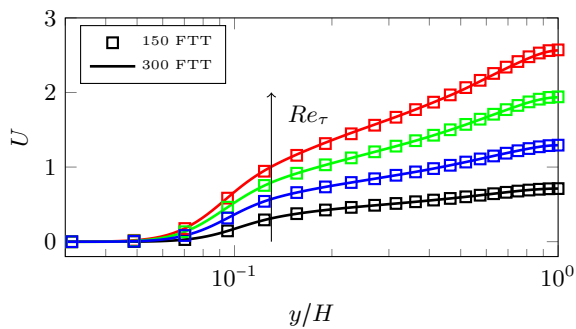}
         \caption{Statistical convergence of minimal channel M1s, color indicates different Re$_\tau=250$, 500, 750, 1000. Lines: 300 FTT, squares: 150 FTT}
         \label{Convtest}
     \end{figure}
 \section*{Appendix B. Convergence test of integration time}
 {In order to determine the appropriate integration time for converged mean velocity profile, mean velocity profiles over roughness $G24M1$ at 4 different Re$_\tau$ averaged with 150 and 300 FTTs are compared and shown in figure 22. The profiles obtained with 150 FTTs at 4 different Re$_\tau$ collapse to the 300 FTTs profiles.
  On the other hand,~\citet{macdonald_chung_hutchins_chan_ooi_garcia-mayoral_2017} estimated the convergence of mean velocity profile in minimal channels in terms of the count of captured $y_c$-sized eddies during the simulation. The number of captured eddies can be expressed as
 \begin{equation*}
         C^*=\frac{T_{sim}u_\tau}{6y_c}\frac{L_y}{H}\frac{L_x}{7.5y_c}\frac{L_z}{2.5y_c},
     \end{equation*} 
     Where $T_{sim}$ is the total simulation time, $y_c=0.4L_z$ is the critical height of the minimal channels. The 95\% confidence interval of the $\Delta U^+$ prediction is formulated as $\Delta U^+\pm\epsilon^+$ where: $\epsilon^+\approx91.4(C^*)^{-1/2}/y_c^+$. The current criteria of 300 FTTs corresponds to $\epsilon^+\approx0.07$, 0.07, 0.05 and 0.03 for minimal channel $M1$ at $Re_\tau=250$, 500, 750 and 1000, respectively.
     Thus the integration time of minimum 300 FTTs in the present work is shown long enough for achieving converged mean velocity profile.}

 \section*{Appendix C. Definition of zero-plane displacement}
 {
       The zero-plane displacement $d$, according to~\citet{jackson_1981}, is placed at the centroid of the distributed drag on the roughness surface. The moment of the drag can be calculated by projecting the drag forces on a y-z plane.
       However, it is demonstrated by this author that $d$ can be calculated in terms of mean flow properties, i.e. total shear stress $\tau_{tot}$. The time averaged Navier-Stokes equation of the flow in streamwise direction writes:
       \begin{equation}
          \rho\frac{\partial(\bar{u}\bar{u})}{\partial x}+\rho\frac{\partial(\bar{u}\bar{v})}{\partial y}+\rho\frac{\partial(\bar{u}\bar{w})}{\partial z}=-\frac{\partial \bar{p}}{\partial x}+\frac{\partial T_{11}}{\partial x}+\frac{\partial T_{12}}{\partial y}+\frac{\partial T_{13}}{\partial z}-P_x+\bar{f}_{x,\text{IBM}}~,
          \label{NS_jackson}
       \end{equation}
       where $\bar{u}$, $\bar{v}$ and $\bar{w}$ are the mean velocity components. $\bar{p}$ is the mean pressure and $T_{11}$, $T_{12}$ and $T_{13}$ are the stresses including Reynolds stresses. $\bar{f_{x,\text{IBM}}}$ is the streamwise component of the mean IBM force. $P_x$ is the constant pressure gradient added to the flow. With the idealized geometry proposed by~\citet{jackson_1981}, if (\ref{NS_jackson}) is integrated over wall-parallel directions, then multiplied by $y$ and integrated over $y$ from the bottom $y=0$ to the tip of the roughness $y=k_t$ we obtain:
       \begin{equation}
           \begin{split}
 &\underbrace{\int_{L_z}\int_0^{k_t}y\rho[\bar{u}\bar{u}]^{L_x}_0dydz}_{=0}+\int_{L_z}\int_{L_x}[y\rho \bar{u}\bar{v}]^{k_t}_0dxdz-\int_{L_z}\int_{L_x}\int^{k_t}_0\rho \bar{u}\bar{v}dydxdz\\
 &+\underbrace{\int_{L_x}\int_0^{k_t}y\rho[\bar{u}\bar{w}]^{L_z}_0dydx}_{=0}=\underbrace{\int_{L_z}\int^{k_t}_0-y[\bar{p}]^{L_x}_0dydz}_{=0}+\underbrace{\int_{L_z}\int_0^{k_t}y[T_{11}]^{L_x}_0dydz}_{=0}\\
 &+\int_{L_z}\int_{L_x}[yT_{12}]^{k_t}_0dxdz-\int_{L_z}\int_{L_x}\int^{k_t}_0T_{12}dydxdz+\underbrace{\int_{L_x}\int_0^{k_t}y[T_{13}]^{L_z}_0dydx}_{=0}\\&-\int_{L_z}\int_{L_x}\int^{k_t}_0yP_xdydxdz+\int_{L_z}\int_{L_x}\int^{k_t}_0y\bar{f}_{x,\text{IBM}}dydxdz~.
 \end{split}
       \end{equation}    
       As marked in the equation, some of the terms vanish due to the periodic boundary condition in wall parallel directions. Thus, the moment of the drag acting on the roughness writes:
       \begin{equation}
     \begin{split}
  M=-\int_{L_z}\int_{L_x}\int^{k_t}_0y\bar{f}_{x,\text{IBM}}dydxdz=\int_{L_x}\int_{L_z}[yT_{12}-y\rho \bar{u}&\bar{v}]_{y=k_t}dzdx\\
  -\int_{L_x}\int_{L_z}\int^{k_t}_0[T_{12}-\rho \bar{u}\bar{v}]dydzdx&-\int_{L_x}\int_{L_z}\int^{k_t}_0yP_xdydzdx~.
 \end{split}
       \end{equation}
       Here $M$ is the moment on the surface.
       Following which, zero-plane displacement $d=k_t-M/(\tau_wL_xL_z)$ is calculated. With the operation $(L_xL_z)^{-1}\int_{L_x}\int_{L_z}[T_{12}-\rho \bar{u}\bar{v}]dzdx$ dispersive stress is included in the total shear stress and is labeled as $\tau_{tot}$~\citep{jackson_1981}.
               \citet{Kameda2018} calculated the displacement $d$ by setting the wall coordinate origin at $k_t$, thus the equation for the zero-plane displacement $d$ writes:
       \begin{equation}
           d=k_t-\frac{\int^{k_t}_0(\tau_{tot}+yP_x)dy}{\tau_w}~.
       \end{equation}}

\bibliographystyle{jfm}
\bibliography{aipsamp,roughness}

\end{document}